\documentclass[fleqn,usenatbib]{mnras}

\usepackage{newtxtext,newtxmath}

\usepackage[T1]{fontenc}
\usepackage{ae,aecompl}
\usepackage{multirow}
\usepackage{tabularx}
\usepackage{supertabular}
\usepackage{longtable}
\usepackage[normalem]{ulem}

\defcitealias{LVK_injection_set}{LVK~\citeyearpar{LVK_injection_set}}

\usepackage[]{url}
\usepackage{graphicx}	

\usepackage{multirow}
\usepackage[left]{lineno}
\usepackage{booktabs}
\usepackage{gensymb}
\usepackage{color}
\usepackage[dvipsnames]{xcolor}

\usepackage[caption = false]{subfig}

\newcommand{\nn}{\nonumber}
\newcommand{\beq}{\begin{equation}}
\newcommand{\eeq}{\end{equation}}
\newcommand{\bdm}{\begin{displaymath}}
\newcommand{\edm}{\end{displaymath}}

\newcommand{\pvtwo}{\texttt{IMRPhenomPv2}}

\newcommand{\pastro}{p_{\rm astro}}
\newcommand{\pastroi}{p_{{\rm astro},i}}
\newcommand{\pastroyj}{p_{{\rm astro}, \gamma(j)}}

\newcommand{\mcdet}{\mathcal{M}_{\rm c,det}}
\newcommand{\chieff}{\chi_{\rm eff}}
\newcommand{\dynesty}{\textsc{dynesty}}
\newcommand{\etabar}{\overline{\eta}}

\definecolor{Gray}{gray}{0.9}
\definecolor{orange}{rgb}{0.9,0.5,0}

\graphicspath{{./plots/}}

\title[Inferring Population of GWs in Presence of Noise]{Inferring the Astrophysical Population of Gravitational Wave Sources in the Presence of Noise Transients}

\author[J.~Heinzel~et~al.]{
Jack~Heinzel$^{1,2}$,
Colm~Talbot$^{1,2}$, 
Gregory~Ashton$^{3}$,
Salvatore~Vitale$^{1,2}$\\
${}^1$ LIGO, Massachusetts Institute of Technology, 77 Massachusetts Avenue, Cambridge, MA 02139, USA \\
${}^2$ Kavli Institute for Astrophysics and Space Research and Department of Physics, \\ Massachusetts Institute of Technology, 77 Massachusetts Avenue, Cambridge, MA 02139, USA \\
${}^3$ Department of Physics, Royal Holloway, University of London, TW20 0EX, United Kingdom, 
}

\begin{document}
\maketitle

\begin{abstract}
The global network of interferometric gravitational wave (GW) observatories (LIGO, Virgo, KAGRA) has detected and characterized nearly 100 mergers of binary compact objects. However, many more real GWs are lurking sub-threshold, which need to be sifted from terrestrial-origin noise triggers (known as glitches). Because glitches are not due to astrophysical phenomena, inference on the glitch under the assumption it has an astrophysical source (e.g. binary black hole coalescence) results in source parameters that are inconsistent with what is known about the astrophysical population. In this work, we show how one can extract unbiased population constraints from a catalog of both real GW events and glitch contaminants by performing Bayesian inference on their source populations simultaneously. In this paper, we assume glitches come from a specific class with a well-characterized effective population (blip glitches). We also calculate posteriors on the probability of each event in the catalog belonging to the astrophysical or glitch class, and obtain posteriors on the number of astrophysical events in the catalog, finding it to be consistent with the actual number of events included. 
\end{abstract}


\section{Introduction}

Since the first direct detection of gravitational waves (GWs) from the merger of two stellar mass black holes \citep{2016PhRvL.116f1102A}, the LIGO-Virgo-KAGRA (LVK) network has observed a large population of these stellar mass binary black holes (BBHs) \citep{2019PhRvX...9c1040A,2021arXiv210801045T,2021arXiv211103606T}. With so many detections comes the ability to characterize the population of BBHs, and shed light on the dominant formation channels of stellar mass BBH mergers. While there is no theoretical consensus on the dominant formation channel, there are many proposals. 

For instance, isolated binary evolution through a common envelope phase \citep{1976IAUS...73...35V, 1976ApJ...207..574S,1993MNRAS.260..675T, 2013A&ARv..21...59I}, stable mass transfer \citep{2017MNRAS.471.4256V}, dynamical many-body interactions in dense stellar environments (e.g. globular clusters, \citealp{1993Natur.364..423S,1993Natur.364..421K,2000ApJ...528L..17P}), chemically homogeneous stellar evolution \citep{2016A&A...588A..50M, 2016MNRAS.458.2634M}, dynamical triples assisted by the Kozai-Lidov mechanism \citep{2017ApJ...841...77A, 2017ApJ...836...39S}, or primordial binary black hole systems \citep{2016PhRvL.116t1301B,2017PhRvD..96l3523A} have been proposed. Traces of these different formation channels are imprinted in the population, distinguishing the relative rates and constraining the sub-population distributions \citep{2021ApJ...910..152Z, 2021hgwa.bookE..16M, 2022LRR....25....1M}. As more GWs are detected, the different astrophysical formation channels will begin to reveal themselves. 

However, one is never sure of the origin of a potential GW detection. GWs are detected using search pipelines, which vary in their methodology, but in general scan the LVK data stream for matches to a GW template within some template bank dense over the expected source parameters \citep{2005PhRvD..71f2001A,2016CQGra..33u5004U,2017ApJ...849..118N,2017PhRvD..95d2001M,2020PhRvD.101b2003H}. This provides a point estimate on the source parameters with the best match template. If this best match passes some significance threshold, it is called a trigger.

GW interferometers are plagued by transient noise fluctuations (known as glitches), whose morphology occasionally mimics real events \citep{2019CQGra..36o5010C, 2017CQGra..34f4003Z, 2021CQGra..38m5014D,2021CQGra..38s5016S,2022arXiv221015633A,2022arXiv221015634A, 2021PTEP.2021eA102A, 2022CQGra..39q5004A}. Most pipelines estimate the false alarm rate (FAR) of a trigger by time-sliding the data of different interferometers by more than the light-travel time between them. Any coincident triggers therefore cannot be caused by a GW propagating at the speed of light, and are deemed false alarms. By varying the time-slide and counting the total number of false alarms, pipelines can accurately estimate the false alarm rate of a trigger. Comparing the FAR to the expected astrophysical rate of the trigger, search pipelines estimate the probability of astrophysical origin, or $\pastro$. In order to calculate the expected astrophysical rate of the trigger, pipelines must assume a model for the underlying astrophysical source population \citep{2021arXiv211103606T}. 

To mitigate contamination from glitches, it is standard to use only the most significant events. Because $\pastro$ estimates assume a population, it is unusual to use pipeline calculated $\pastro$ as a threshold for population inference. Instead, a common threshold is FAR $< 1$yr$^{-1}$, yet even with this high threshold, one expects e.g. 4.6 false alarms in the catalog used by \cite{PhysRevX.13.011048} under the assumption that the search pipelines produce events independently \citep{2005PhRvD..71f2001A,2016CQGra..33u5004U,2017ApJ...849..118N,2017PhRvD..95d2001M,2020PhRvD.101b2003H,PhysRevX.13.011048}. 
Therefore, one must tune the FAR threshold to minimize the systematic uncertainty of including more false alarms in the catalog, and the statistical uncertainty of including fewer events. 

There are also a plethora of sub-threshold (FAR > $1 {\rm yr}^{-1}$) astrophysical events which contain information about the population of gravitational-wave sources in the Universe, especially in some of the more poorly measured regions of parameter space, where glitches are responsible for reduced search sensitivity. Sub-threshold mergers of binary neutron stars (BNS), neutron star black holes (NSBH) or stellar mass BBHs can improve known constraints on the population of these as gravitational wave progenitors. Indeed, there are many more events with lower significance; the rate of GW events scales with SNR${}^{-4}$, assuming a constant merger rate in a Euclidean volume \citep{2011CQGra..28l5023S, 2014arXiv1409.0522C}. Though these lower significance events also encode less information about the progenitor, events as low as SNR$\sim 6-7$ can have well-measured chirp masses \citep{2018PhRvD..98l3021H}. 

Moreover, certain kinds of theoretical GW events may pass this FAR threshold only rarely, with the majority falling deep into the sub-threshold range. For instance, subsolar-mass compact objects are predicted by certain modifications to the standard model of particle physics or $\Lambda$CDM \citep{2018PhRvL.120x1102S,2021ApJ...915...54N,2023MNRAS.tmp..611A}. Though no direct detections have been made of a sub-solar mass merger \citep{2019PhRvL.123p1102A, 2021ApJ...915...54N}, it is possible there are some lurking within the large set of sub-threshold candidates; because of their low masses, the signal-to-noise ratio (SNR) and significance of the GW will be much lower.

Glitches in GW interferometers are commonly studied by modelling the data as some parametric and deterministic function plus a stationary and stochastic noise process \citep{2015CQGra..32m5012C, 2021PhRvD.104j2004M, 2023arXiv230110491T,2023ApPhL.122i4103U}. This is preferable to modelling glitches as some general non-stationary noisy time series, where the statistical properties are unclear. A glitch model then requires a parametric function, called the glitch waveform, for the deterministic part of the signal. Since significant false alarms will mimic real GWs, it is sensible to use a GW model for the glitch waveform. In this paper, we follow this prescription, modelling glitches with a GW waveform. 

A more general glitch model distinguishes GWs from terrestrial glitches by signal coherence. Real GWs must be coherent between multiple detectors and the waveforms should be consistent with the same progenitor parameters, while the same is not true for coincident false alarms \citep{2010PhRvD..81f2003V}. Glitches may therefore be modelled as an independent GW waveform in each detector, relaxing this coherence requirement. This is justified as a worst-case scenario, where a background event is distinguished from an astrophysical one based purely on the signal coherence. This glitch model has been used to calculate the probability an event is astrophysical \citep{2018PhRvD..98d2007I, 2019PhRvD.100l3018A,2021PhRvD.104l4039P}, and to rule out marginal candidates (e.g. \citealt{2020MNRAS.498.1905A, 2022MNRAS.516.5309V}). 
The most general glitch models make no physical assumptions about the source and model glitches as a superposition of wavelets \citep{2015CQGra..32m5012C}.

Whatever the waveform assumed for the glitches, a population would then be given by probability distributions on their parameters. Indeed, it is possible to study the population of glitches and astrophysical events simultaneously, allowing for each event to belong to either class. Previous work approached this problem from different perspectives. \cite{2015PhRvD..91b3005F} showed how to infer the rates of astrophysical and background populations when the shapes of the populations are known, but the identity of each event (i.e. which population it originates from) is unknown.  \cite{2019MNRAS.484.4008G} show that it is indeed possible to do joint inference on an astrophysical and a glitch population, but leave a study with real GW data for a future analysis.  \cite{2020PhRvD.102l3022R, 2020PhRvD.102h3026G} analyze real GW data, and fold in pipeline information -- in particular, $\pastro$ estimates, to build a glitch population model. However, this carries a fixed background event rate estimate by each search pipeline, rather than inferring the rate of events from the background population in a Bayesian manner. 

In this paper, we present a general method to simultaneously model the population of background non-astrophysical triggers and the population of astrophysical objects, in a fully Bayesian manner. We use a population of short glitches (``blips'') as identified by the \textsc{GravitySpy} algorithm \citep{2017CQGra..34f4003Z} to contaminate the catalog of astrophysical signals. While this is done for computational expedience, the method can be used for any type of non-astrophysical transients, as long as one can characterize their ``usual'' properties. Similarly. while we focus on the population of BBHs, the method may be used to study any population of foreground events contaminated with undesirable background events. In section \ref{sec:methods}, we briefly review Bayesian parameter estimation of GW sources and population inference. Then, we discuss how this picture is complicated when one allows for the possibility that the dataset is contaminated by glitches. In section \ref{sec:glitchpop} we discuss our glitch population parameterization and constrain the population hyperparameters using a large representative sample. In section \ref{sec:results}, we contaminate a catalog of GWs with glitches, and show how our method consistently models and removes the bias due to the contaminants. Finally, in section \ref{sec:conclusion}, we summarize and discuss future work.

\section{Methods}
\label{sec:methods}

\subsection{Parameter Estimation}

Consider a stretch of LVK frequency domain data $d$ which is a sum of noise $n$ and waveform signal $h(\theta)$
\beq
d = h(\theta) + n,
\label{eq:noise}
\eeq
where $\theta$ represents the unknown parameters of the GW source. 
Approximating the noise as stationary and gaussian, the likelihood can be written 
\beq
\log\mathcal{L}(d|\theta) = -\sum_j\left(2\Delta f\frac{|d_j - h_j(\theta)|^2}{\mathscr{P}_j} + \log(2\pi\mathscr{P}_j)\right),
\label{eq:PE_like}
\eeq
where $d_j$ and $h_j$ represent the $j^{\rm th}$ frequency component of the data and waveform, respectively, $\mathscr{P}_j$ is the power-spectral-density, and $\Delta f$ is the frequency spacing \citep{whittle1951hypothesis}. With this likelihood, a model for the waveform $h(\theta)$ given some GW parameters, and priors for the GW parameters, one can then sample from the posterior of the GW parameters \citep{2015PhRvD..91d2003V,2019PASA...36...10T,2022RvMP...94b5001C}. 

The above process also can apply to glitches, thinking of them as a deterministic signal buried in stochastic noise. Modelling glitches under some parameterization (e.g. a sine-gaussian), one can perform parameter estimation exactly as above for the glitch parameters, which we denote $\psi$. Indeed, while glitches are usually ruled out by search pipelines by e.g. $\chi^2$ discriminators \citep{2005PhRvD..71f2001A}, there can be cases where glitches are mistaken for astrophysical GWs. Because population inferences generally assume that all events in the catalog are truly astrophysical, a contaminant glitch in the catalog will bias the inference. We want to relax this assumption, and jointly infer the population of astrophysical events and glitches. 

\subsection{Population Inference Without Glitches}

Before we discuss simultaneous inference of the astrophysical and glitch populations, we review the general GW population inference problem. Given posterior samples from a set of data timeseries $\{d_i\}_{1\le i\le N_{\rm events}}$, one can write the likelihood for a population model. In general, a population model describes the rate of mergers within a small interval of GW parameter space $[\theta, \theta + d\theta]$. However, the rate is typically assumed to be a Poisson process, and we can instead write down a probability density $p_A(\theta|\Lambda)$, irrespective of the overall rate. Here, $\Lambda$ are called the hyper-parameters; a finite list of parameters which vary the shape of the population distribution (e.g. the mean and variance of a gaussian, the power index to a power-law, etc.). We give the subscript $A$ to refer to ``astrophysical.'' This is in contrast to $G$ for ``glitch,'' which we will use later in this paper. 

Assuming a Poisson process for the events and marginalizing over the overall rate $R$ with an uninformative (uniform in $\log R$) prior, one obtains the hierarchical likelihood
\beq
\mathcal{L}(\{d_i\} | \Lambda) \propto \prod_{i=1}^{N_{\rm events}} \frac{\int d\theta \mathcal{L}(d_i|\theta)p_A(\theta|\Lambda)}{\alpha(\Lambda)}
\label{eq:hyper-like}
\eeq
and the selection function
\beq 
\alpha(\Lambda) = \int d\theta p_{{\rm det},A}(\theta)p_A(\theta|\Lambda)
\label{eq:alpha}
\eeq 
is the fraction of events which are detectable in the population with hyperparameters $\Lambda$ (for a derivation of the likelihood see \citealt{2019MNRAS.486.1086M,Vitale2020}). The quantity $p_{{\rm det},A}(\theta)$ is the probability of detecting an astrophysical event with parameters $\theta$, given by
\beq
p_{{\rm det},A}(\theta) = \int_{\{d \in \mathcal{D} |\rho(d) > \rho_{\rm thr}\}} \mathcal{L}(d | \theta) {\rm d}d
\label{eq:pdet}
\eeq
the integral over all possible data realizations which exceed the detection threshold $\rho(d) > \rho_{\rm thr}$ (i.e. FAR < $1{\rm yr}^{-1}$, as in \citealt{PhysRevX.13.011048}). 

In practice, the integrals in Eq. \ref{eq:hyper-like} and \ref{eq:alpha} are estimated with Monte Carlo estimators. In particular, 
\beq
\int d\theta \mathcal{L}(d_i|\theta)p_A(\theta|\Lambda) \sim \frac{Z(d_i)}{N_{\rm samp}}\sum_{j=1}^{N_{\rm samp}}\frac{p_A(\theta_j | \Lambda)}{\pi(\theta_j|\mathcal{H}_{\rm PE})}\bigg|_{\theta_j \sim p(\theta | d_i)},
\eeq
where $\theta_j$ are samples from the $i^{\rm th}$ event posterior, 
\beq
Z(d_i) = \int d\theta \mathcal{L}(d_i|\theta)\pi(\theta |\mathcal{H}_{\rm PE})
\eeq
is the evidence and $\pi(\theta|\mathcal{H}_{\rm PE})$ is the sampling prior used for the parameter estimation. 
As for the selection function,
\beq
\alpha(\Lambda) \sim \frac{1}{N_{\rm draw}}\sum_{j=1}^{N_{\rm det}}\frac{p_A(\theta_j|\Lambda)}{p_{\rm draw}(\theta_j)}\bigg|_{\theta_j \sim p_{\rm draw}(\theta)},
\eeq
where $N_{\rm draw}$ events are drawn from some fiducial distribution $p_{\rm draw}(\theta)$, data drawn from the conditioned likelihood $\mathcal{L}(d|\theta)$ with a suitable power spectral density choice, and then search pipelines run to recover $N_{\rm det}$ of the total events (for details see e.g. \citealt{2018CQGra..35n5009T,2019RNAAS...3...66F}).

\subsection{Population Inference With Glitches}

The above procedure assumes every event which passes the threshold is a real GW. This assumption can be relaxed by simultaneously fitting the glitch population. Suppose the glitch waveform is given by parameters $\psi$, and we obtain posteriors on $p(\psi | d_i)$ for each event in the catalog, as well as posteriors on $p(\theta|d_i)$ for the GW parameters. With Eq. 79 in \cite{Vitale2020} and a relative rate $\eta$ of GWs versus a GW-like glitches, one can marginalize over the total rate with a uniform in $\log R$ prior to generalize Eq. \ref{eq:hyper-like}.
\begin{align}
&\mathcal{L}(\{d_i\} | \Lambda_A, \Lambda_G, \eta) \propto  \nn \\ &\prod_{i=1}^{N_{\rm events}} \frac{\eta \int d\theta \mathcal{L}(d_i|\theta)p_A(\theta|\Lambda_A) + (1-\eta) \int d\psi \mathcal{L}(d_i|\psi)p_G(\psi|\Lambda_G)}{\eta \alpha_A(\Lambda_A) + (1-\eta) \alpha_G(\Lambda_G)},
\label{eq:hyper-like-glitch}    
\end{align}
where $\Lambda_A$ and $\Lambda_G$ refer to the astrophysical and glitch hyperparameters, $p_G(\psi|\Lambda_G)$ is the population model for the glitch waveform parameters and $\alpha_X(\Lambda_X)$ is the selection function for the $X$ subpopulation:
\beq
\alpha_X(\Lambda_X) = \int d\theta p_{{\rm det},X}(\theta)p_X(\theta|\Lambda_X)
\eeq
$p_{{\rm det},G}$ is analogous to the $p_{{\rm det},A}$ we defined above, but we want to allow for the possibility that the detection criterion $\rho(d) > \rho_{\rm thr}$ is different for glitches. In reality, the same detection criterion must be used for all events for a catalog, but for reasons we will describe below, we must use a different detection criterion for glitches in this study.

The mixing fraction $\eta$ represents the relative rate of all GWs from all GW-like sources (astrophysical and glitches), whether they are detected or not. It is useful to define a detectable mixing fraction:
\beq
\etabar = \frac{\eta \alpha_A(\Lambda_A)}{\eta \alpha_A(\Lambda_A) + (1-\eta)\alpha_G(\Lambda_G)}
\eeq
which is the fraction of detectable events which are GWs.
In this case, and a bit of algebra, the likelihood of Eq. \ref{eq:hyper-like-glitch} can be recast as 
\begin{align}
&\mathcal{L}(\{d_i\} | \Lambda_A, \Lambda_G, \eta) \propto  \nn \\ &\prod_{i=1}^{N_{\rm events}} \frac{\etabar \int d\theta \mathcal{L}(d_i|\theta)p_A(\theta|\Lambda_A)}{\alpha_A(\Lambda_A)} + \frac{(1-\etabar) \int d\psi \mathcal{L}(d_i|\psi)p_G(\psi|\Lambda_G)}{\alpha_G(\Lambda_G)},
\label{eq:hyper-like-glitch-alt}    
\end{align}
which is the form of the likelihood we will use in the sampling.

So far we have assumed glitches and GWs will be characterized with different parameters, $\theta$ and $\psi$. However, glitches which can contaminate a GW catalog will necessarily be well modelled by a GW waveform. For this proof-of-principle analysis, we thus model the waveform of a glitch as a GW (we set $\psi \to \theta$). Furthermore, we only model the population in the intrinsic GW parameters; this will be explained further in section \ref{sec:glitchpop}. This simplifies the analysis: we don't need evidences and posterior samples for every event under both the glitch and GW hypotheses--both analyses are the same. Indeed, under these assumptions the analysis reduces to a GW population inference with a mixture population; Eq. \ref{eq:hyper-like-glitch} becomes Eq. \ref{eq:hyper-like} with 
\beq
p(\theta | \Lambda) \to \eta p_A(\theta | \Lambda_A) + (1-\eta) p_G(\theta | \Lambda_G),
\label{eq:mixture_model}
\eeq
and a selection function 
\beq
\alpha(\Lambda) \to \; \eta \alpha_A(\Lambda_A) + (1-\eta) \alpha_G(\Lambda_G).
\label{eq:alpha_mix}
\eeq
Eq. \ref{eq:mixture_model} treats the glitch population as an additional ``astrophysical'' population, albeit occupying a different region of parameter space from the population of true astrophysical BBHs. 

There is one additional caveat. In the LVK population analysis of \cite{PhysRevX.13.011048}, events included in the catalog are selected by their FAR ($< 1 {\rm yr}^{-1}$), and so we would like to also select glitches by their FAR to match \cite{PhysRevX.13.011048}. However, this requires us to calculate FARs for many injections from a fiducial glitch population. Running search pipelines to calculate FARs of injected glitches may be necessary for a future study, however for this proof-of-principle paper it is simply too expensive. Instead, we select glitches for inclusion with a cheaper threshold, the signal-to-noise ratio (SNR). We can then estimate $\alpha_G(\Lambda_G)$ with a reweighted Monte Carlo estimator using a custom set of injections, and estimate $\alpha_A(\Lambda_A)$ with the injection set already provided in \citetalias{LVK_injection_set}. 

\subsection{Characterizing the Glitch Population}
\label{sec:glitchpop}
\begin{figure}
    \centering
    \includegraphics[width=0.85\linewidth]{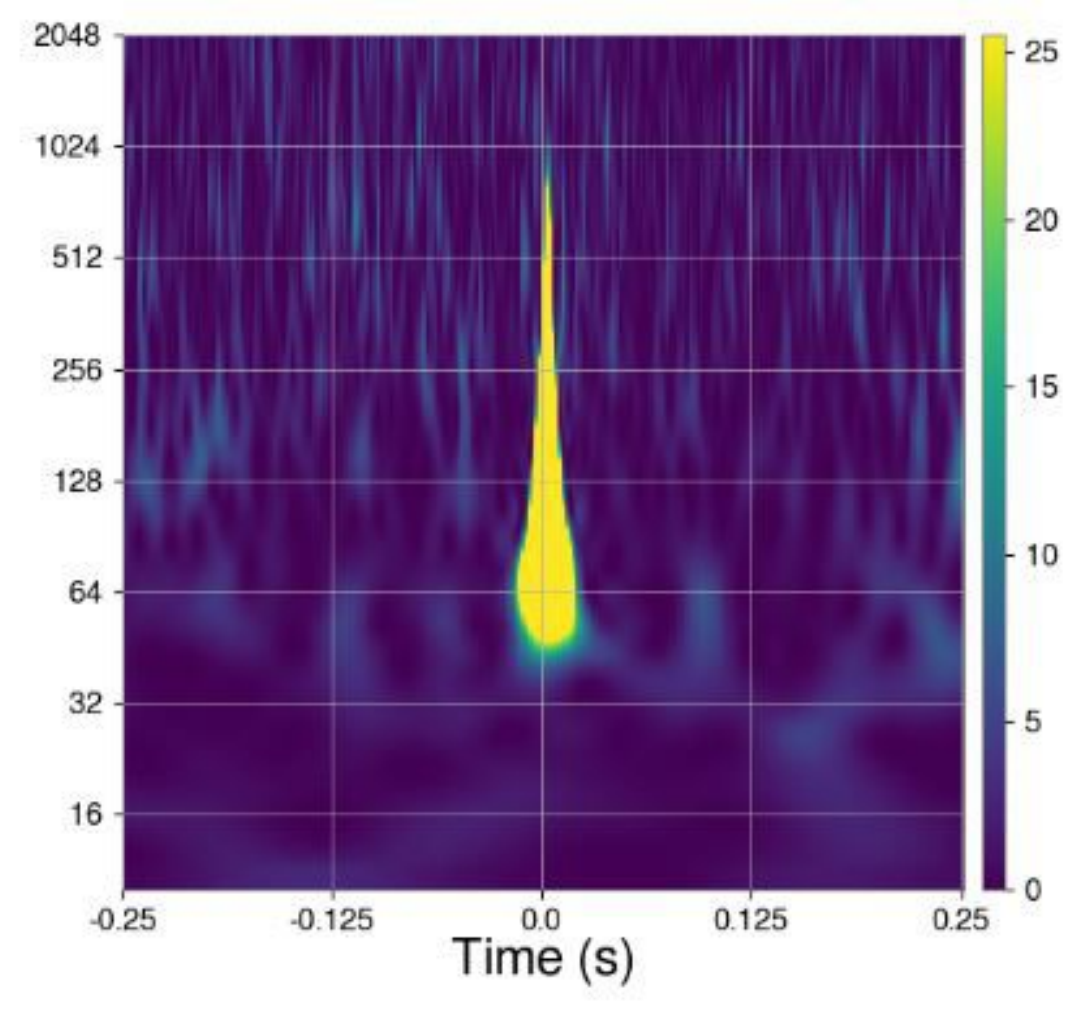}
    \caption{A time frequency spectrogram of a \textsc{GravitySpy}-identified blip glitch in LIGO-Hanford. This blip occurred on August 28, 2019 at UTC 16:56:49. 
    Due to their short duration, blip glitches can be mistaken for high mass BBHs.}
    \label{fig:blip}
\end{figure}

In the citizen-science project \textsc{GravitySpy}, glitches are classified according to their time frequency spectrograms \citep{2017CQGra..34f4003Z, 2023CQGra..40f5004G}. For instance, blip glitches are short bursts of excess power, with a time frequency spectrogram morphology shown in Fig. \ref{fig:blip}.
In fact, blip glitches are more likely to contaminate a GW catalog, since they can mimic high mass BBHs \citep{2019CQGra..36o5010C}. For this reason, we restrict this first study to blip glitches, though the formalism can be extended to any glitch class, or even combination of classes. This would require a new population model and $\Lambda_G$ for each additional class, plus a mixing fraction.

In order to understand various populations of glitches, \cite{2022CQGra..39q5004A} analyzed a set of 1000 \textsc{GravitySpy} identified blip glitches with the \pvtwo~ GW waveform \cite{HannamIMRPhenomP,Bohe2016,IMRPhenomPv2-I_husa2016,IMRPhenomPv2-II_khan2016}. Since blip glitches are not due to any astrophysical process, they are usually present in a single detector, with multiple detector coincidences occurring randomly. As single detector triggers, only information about the intrinsic parameters (masses and spins) may be extracted. Therefore \cite{2022CQGra..39q5004A} provides posterior samples only over the intrinsic parameters and the redshift. 

With the posterior samples in hand, \cite{2022CQGra..39q5004A} fit a population model in the detector frame chirp mass, mass ratio, and primary spin (see their Fig. 2, 3 \& 4). Qualitatively, the population of \textsc{GravitySpy} blip glitches shows different features from the population of BBHs (extreme mass ratios, spins, and low redshifts, inconsistent with e.g. \citealt{PhysRevX.13.011048}). We will use this to our advantage to separate the populations. 

We slightly modify the population model of \cite{2022CQGra..39q5004A}. Instead of modelling the primary spin magnitude, we model in the effective spin parameter:
\beq
\chieff = \frac{a_1 \cos \theta_1 + q a_2 \cos\theta_2}{1+q}
\label{eq:chieff}
\eeq
where $a_1$ and $a_2$ are the spin magnitudes of the primary and secondary BHs in Kerr units, $q = m_2/m_1$ is the mass ratio (where $0<q<1$ by convention), and $\theta_1$ and $\theta_2$ are the spin tilts measured from the orbital angular momentum. $\chieff$ is the spin parameter which occurs at lowest order in the waveform, and is measured better than individual spins \citep{2008PhRvD..78d4021R,2016PhRvD..93h4042P,2017PhRvD..95f4053V, 2018PhRvD..98h3007N}. 

We model the glitch population in the detector-frame chirp mass, mass ratio, effective spin parameter, and redshift: $\theta = (\mcdet, q, \chieff, z)$. In particular, we use a skewed gaussian (Eq. \ref{eq:skewgauss}) for both the detector-frame chirp mass $\mcdet$ and the redshift $z$ with hyperparameters $\mu_m, \sigma_m, \kappa_m$ and $\mu_z, \sigma_z, \kappa_z$ respectively
\beq
p(x|\mu, \sigma, \kappa) = \frac{2}{\sigma}\phi\left(\frac{x - \mu}{\sigma}\right)\Phi\left(\kappa\frac{x - \mu}{\sigma}\right),
\label{eq:skewgauss}
\eeq
where $\phi$ and $\Phi$ are the standard gaussian and gaussian integral, respectively.
We model $\chieff$ and $q$ with a correlated mixture model of two two-dimensional gaussians in the $\chieff-q$ plane with hyperparameters denoted $\vec{\lambda}_{q\chi}$ for brevity
\begin{align}
p_{q\chi}(q,\chieff | &\vec{\lambda}_{q\chi}) = N_1 \eta_{q,\chi}\phi\left(f_1[q,\chieff]\right)\phi\left(g_1[q,\chieff]\right) \nn \\
&+ N_2 (1-\eta_{q,\chi})\phi\left(f_2[q,\chieff]\right)\phi\left(g_2[q,\chieff]\right) \\ 
\vec{\lambda}_{q\chi} = (\mu_{q,1}.&\mu_{q,2},\mu_{\chi,1},\mu_{\chi,2},\sigma_{q,1},\sigma_{q,2},\sigma_{\chi,1},\sigma_{\chi,2},\theta_{q,\chi},\eta_{q,\chi}) \nn
\label{eq:chieffq}
\end{align}
where
\begin{align}
f_i[q,\chieff] &= \frac{(q-\mu_{q,i})\cos(\theta_{q,\chi}) + (\chieff - \mu_{\chi,i})\sin(\theta_{q,\chi})}{\sigma_{q,i}} \\ 
g_i[q,\chieff] &= \frac{(q-\mu_{q,i})\sin(\theta_{q,\chi}) + (\chieff - \mu_{\chi,i})\cos(\theta_{q,\chi})}{\sigma_{q,i}},
\end{align}
and $N_1$ and $N_2$ are normalization coefficients, numerically calculated because $\chieff$ and $q$ are required to be positive.
Eq. \ref{eq:chieffq} describes a pair of two dimensional gaussians with branching fraction $\eta_{q,\chi}$, parameterized by the variances along the eigenvectors of the covariance matrix ($\sigma_{q,i}^2$ and $\sigma_{\chi,i}^2$), and the angle they are ``tilted'' by ($\theta_{q,\chi}$, assumed to be the same for both gaussians). 
The glitch population model for is the product of the $\mcdet$, $z$, and $q-\chieff$ models, and $\Lambda_G$ is the union of their hyperparameters. We chose to leave precession unmodeled in the population by projecting the 6-dimensional spin population onto the effective aligned spin parameter. However a future study could examine how the populations further separate including the spin precession parameter and correlations therein, or in the full 6-dimensional spin space.

\begin{figure}
    \centering
    \includegraphics[width=0.9\linewidth]{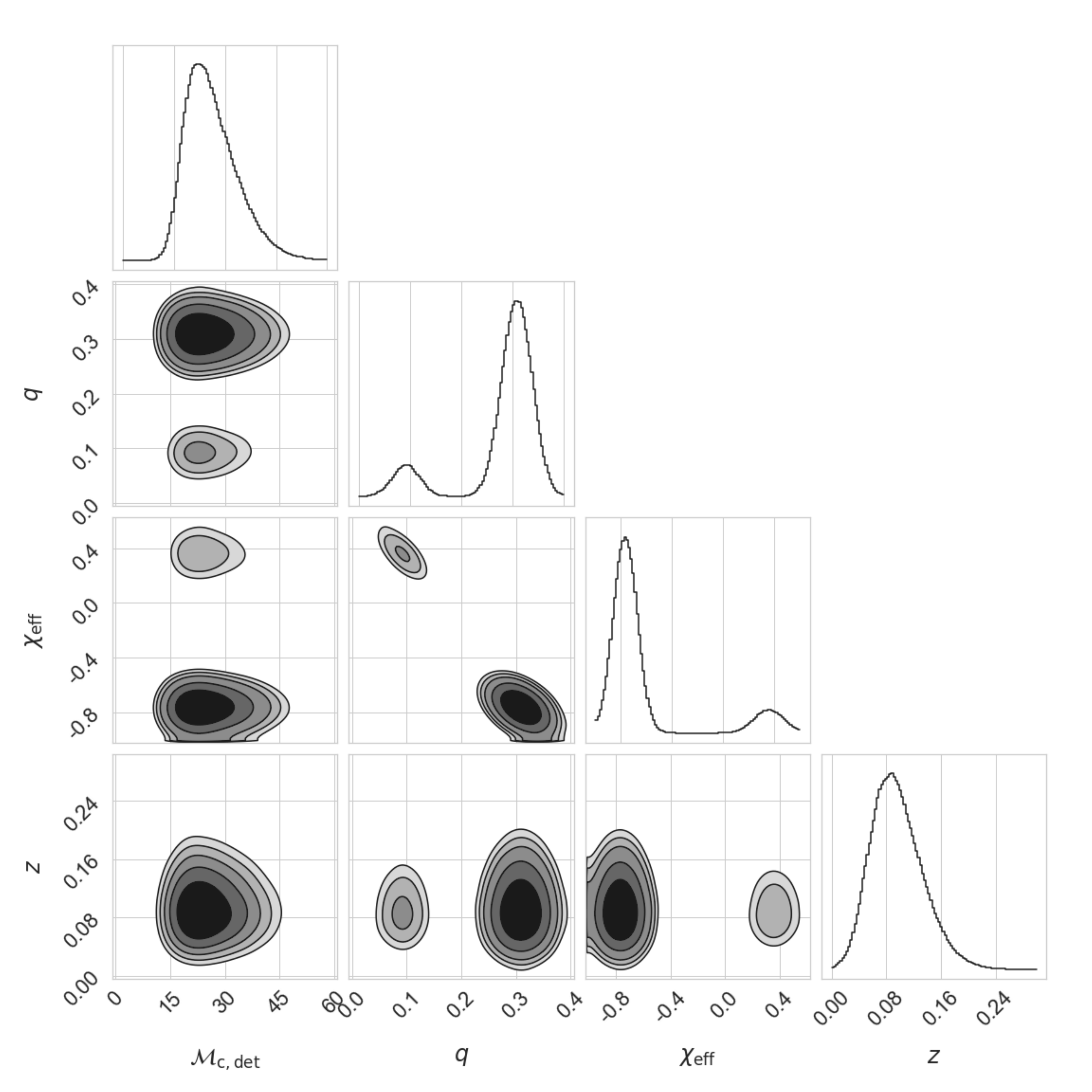}
    \caption{The posterior population distribution of the population inference on the glitch population alone, using all 1000 blip posteriors from \protect\cite{2022CQGra..39q5004A}. Contours show the 1-5$\protect\sigma$ regions.}
    \label{fig:glitch_pop_ppd}
\end{figure}

We are now ready to measure the population of blip glitches with our model, as is done in \cite{2022CQGra..39q5004A}. By using all 1000 posteriors from \cite{2022CQGra..39q5004A} we obtain tight constraints on the glitch population alone. This is a critical step of our analysis. We must measure the population of glitches well to optimally separate it from the population of GWs. Fortunately, we have access to the unbiased population of blip glitches before any selection criteria are enforced.\footnote{Note there is a cut on these \textsc{GravitySpy} glitches with SNR$>8$, and another given they are \textsc{GravitySpy} identified. We can still treat this as the unbiased population with no changes to our analysis. In a real analysis, one would still have access to an unbiased sample of the population of glitches.} We may assume the 1000 glitches from \cite{2022CQGra..39q5004A} are a representative sample. The constraints we measure in this step inform the boundaries of the priors we use during simultaneous inference.
We show the posterior population distribution (PPD) in Fig. \ref{fig:glitch_pop_ppd}.

As for the astrophysical population parameterization, we use the powerlaw plus peak model of \cite{PhysRevX.13.011048, 2018ApJ...856..173T} and the redshift model of \cite{2018ApJ...863L..41F}. We modify the spin distribution model by modelling $\chieff$ with a gaussian, following \cite{2019MNRAS.484.4216R,2020ApJ...895..128M,2021ApJ...922L...5C}. This gives us the set of $\Lambda_A$ and $\Lambda_G$, which will be inferred together with the detectable mixing fraction $\etabar$ in our joint analysis. 

\subsection{Simulatenous Inference and Selection Effects}

We model the selection effects of glitches in entirely the same way we model the selection effects of GWs.
We emphasize that selection effects depend on the data alone. If we believe glitches have data well modelled by a GW plus gaussian noise, then the probability of detecting a glitch is well approximated by the probability of detecting a GW with the corresponding parameters. 

We also define $\pastroi(\Lambda)$ for each event in the catalog, a population dependent quantity, 
\beq
\pastroi(\Lambda) = p({\rm astro} | d_i, \Lambda) = \frac{\eta \mathcal{L}(d_i|\Lambda_A)}{\eta \mathcal{L}(d_i|\Lambda_A) + (1-\eta) \mathcal{L}(d_i|\Lambda_G)}.
\label{eq:pastro}
\eeq
This comes directly from Bayes' Theorem. It is perhaps more intuitive to use $\etabar$ instead of $\eta$, however in that case the likelihood terms must each acquire a $1/\int d\theta p_{{\rm det},X}p_X(\theta|\Lambda_X)$ term, and it reduces again to Eq. \ref{eq:pastro}. This folds in the dependence on source parameters and uncertainty in the population hyperparameters, and so in general $\pastro$ is a posterior, based on the posterior on $\Lambda$. Search pipelines output a point-estimate of this quantity for each event, using the point estimate on the progenitor parameters with the matched template, a fixed underlying astrophysical population, and a direct calculation of the glitch-rate term with the FAR. \cite{2015PhRvD..91b3005F, 2020CQGra..37d5007K} define similar quantities.

\section{Results}
\label{sec:results}
We contaminate the catalog of 69 BBH events (with FAR < $1{\rm yr}^{-1}$) analyzed in \cite{PhysRevX.13.011048} with blip posteriors obtained from \cite{2022CQGra..39q5004A}. Note the posteriors in \cite{PhysRevX.13.011048} are sampled used state-of-the-art waveforms including higher order modes, while \cite{2022CQGra..39q5004A} uses the rapid \pvtwo, a waveform approximant including only the dominant $(l,m)=(2,2)$ mode~\citep{IMRPhenomPv2-I_husa2016,IMRPhenomPv2-II_khan2016,HannamIMRPhenomP,Bohe2016}. Indeed, blip posteriors converge on unequal mass ratios ($q \sim 0.1$), where higher order modes become significant. While this will bias the glitch population model, this paper is intended to be a proof of concept and so we use the posterior samples as provided. 

We inject $N_{\rm blip} = [0,1,2,...,19,20]$ contaminant posteriors from \cite{2022CQGra..39q5004A} into the set of 69 BBH posteriors analyzed in \cite{PhysRevX.13.011048}. We then sample the hyperposterior of $\Lambda$ using the nested sampler $\dynesty$ \citep{2020MNRAS.493.3132S,KoSp2022} and the code \textsc{gwpopulation} \citep{2019PhRvD.100d3030T}. We must cut regions of parameter space above the total variance of the hierarchical likelihood-estimator. Without handling variance the sampler can converge on regions of parameter space with poor Monte Carlo estimates, and thereby bias the posterior sampling from the true posterior \citep{2022arXiv221012287G}. We do this as well as cut out regions with poorly behaved selection function estimates, as described in \cite{2019RNAAS...3...66F,2022arXiv220400461E}. 


We describe several methods of quantifying the bias (or lack thereof) of performing the simultaneous inference.  
\begin{figure}
    \centering
    \includegraphics[width=0.95\linewidth]{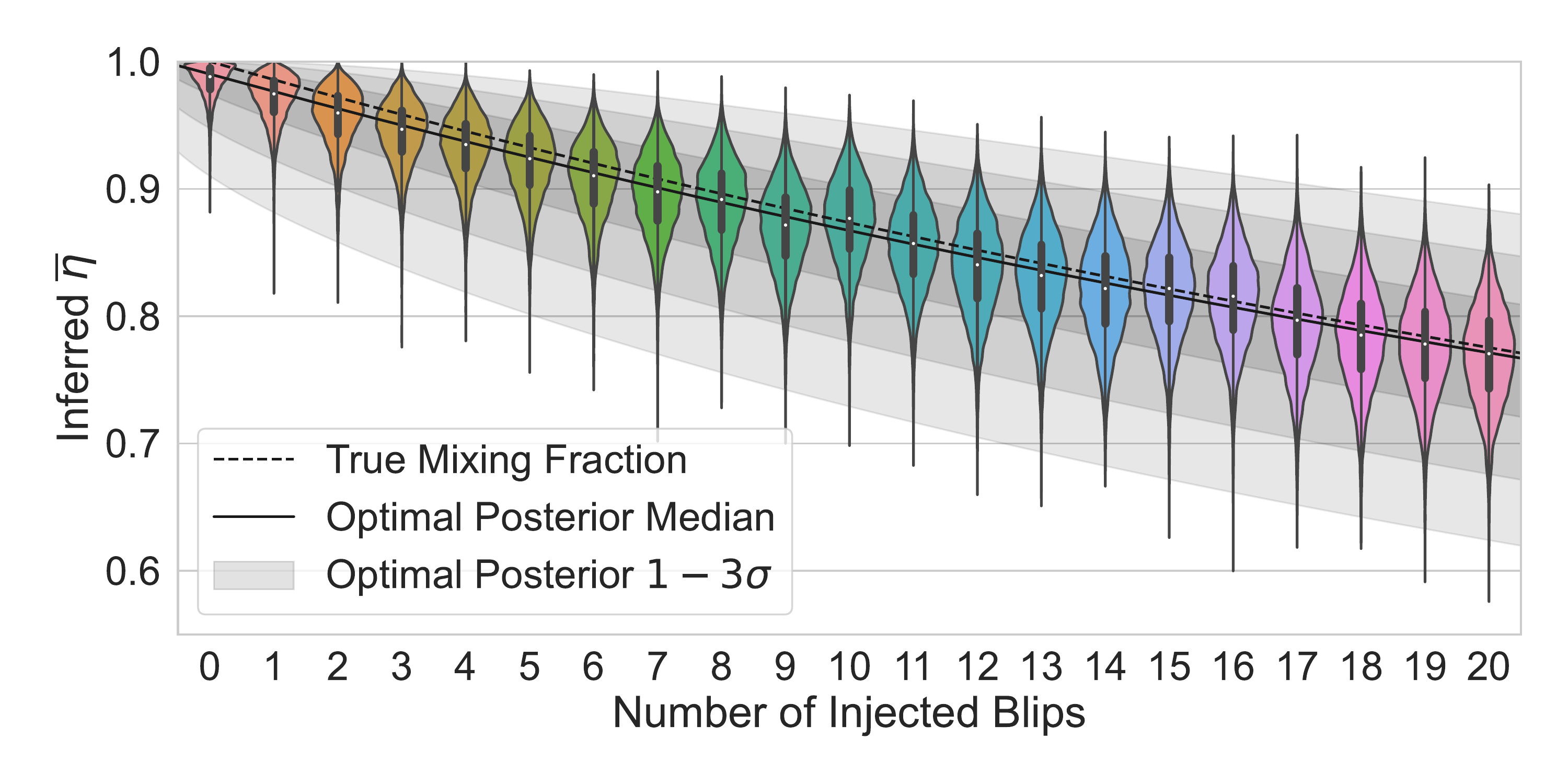}
    \caption{The violins show the inferred detectable mixing fraction $\etabar$ for each run. The $x$-axis indexes the number of injected blips and each violin refers to a different inference. The black dashed line is the injected mixing fraction, given by $1 - N_{\rm blips} / N_{\rm events}$. Notice the inference recovers the injected mixing fraction well. We compare against the optimal posterior which would be inferred with perfect knowledge on which events are BBHs and are glitches. We show the 1-3$\sigma$ and median of this optimal posterior in black (see the appendix and Eq. \ref{eq:beta_model}).}
    \label{fig:etabar}
\end{figure}

\subsection{The Detectable Mixing Fraction}

For each catalog and its inference, we obtain a posterior on the detectable mixing fraction $\etabar$. We plot these posteriors as violin plots in Fig. \ref{fig:etabar}. The dashed black line is the true mixing fraction in our catalog, given by the number of BBHs divided by the total number of events. Note the posteriors peak at the dashed line, i.e. it is recovering the correct number of contaminants.

While the results here suggest the sampling is correctly recovering the blips, there is a caveat. The quantity $\etabar$ represents a statement on the underlying relative rates, it is not the fraction 
of BBHs in the catalog. In other words, this is not a like to like comparison. We want to understand what our inference predicts are the number of BBHs and blips in our catalog. 

For instance, suppose we may unambiguously identify which events are BBHs and which are blips solely using the event parameters. That is, the populations are disjoint to the point that every event posterior overlaps with only one of the astrophysical or glitch populations. It turns out that $\etabar$ does not converge on a delta function: it will have some width due to Poisson rate uncertainty. Rather, it converges on an analytic optimal posterior, which we calculate by assuming the populations are so disjoint that every event posterior uniquely determines which population the event originates from. Details on this calculation are in the appendix.

From this theoretical optimal posterior, we can calculate the median and 1-3$\sigma$ levels, which we show as a function of the number of added contaminants ($x$-axis) in Fig. \ref{fig:etabar}. Note how similar the measured posteriors on $\etabar$ are to the optimal posterior given perfect knowledge on which events are BBHs and glitches. The populations of blip glitches and BBHs are nearly disjoint; this suggests the inference can uncover which events are in which population much more precisely than the $\etabar$ posteriors naively indicate. 

\subsection{Inferred Number of Contaminants and BBHs} 

Calculating $\pastro$\footnote{We emphasize that statements made in this paper about $\pastro$ should be understood as the probability of the event not being a blip, rather than the probability of the event being astrophysical in origin. This is rather cumbersome to write, so we continue with the abuse of notation in $\pastro$.} (i.e. Eq. \ref{eq:pastro}) for each event in each run, we notice that the posteriors on each event tends to be sharply peaked, e.g. GW150914 peaks at $\pastro(\Lambda) \to 1$, the blips peak at $\pastro(\Lambda) \to 0$.
\begin{figure}
    \centering
    \includegraphics[width=0.98\linewidth]{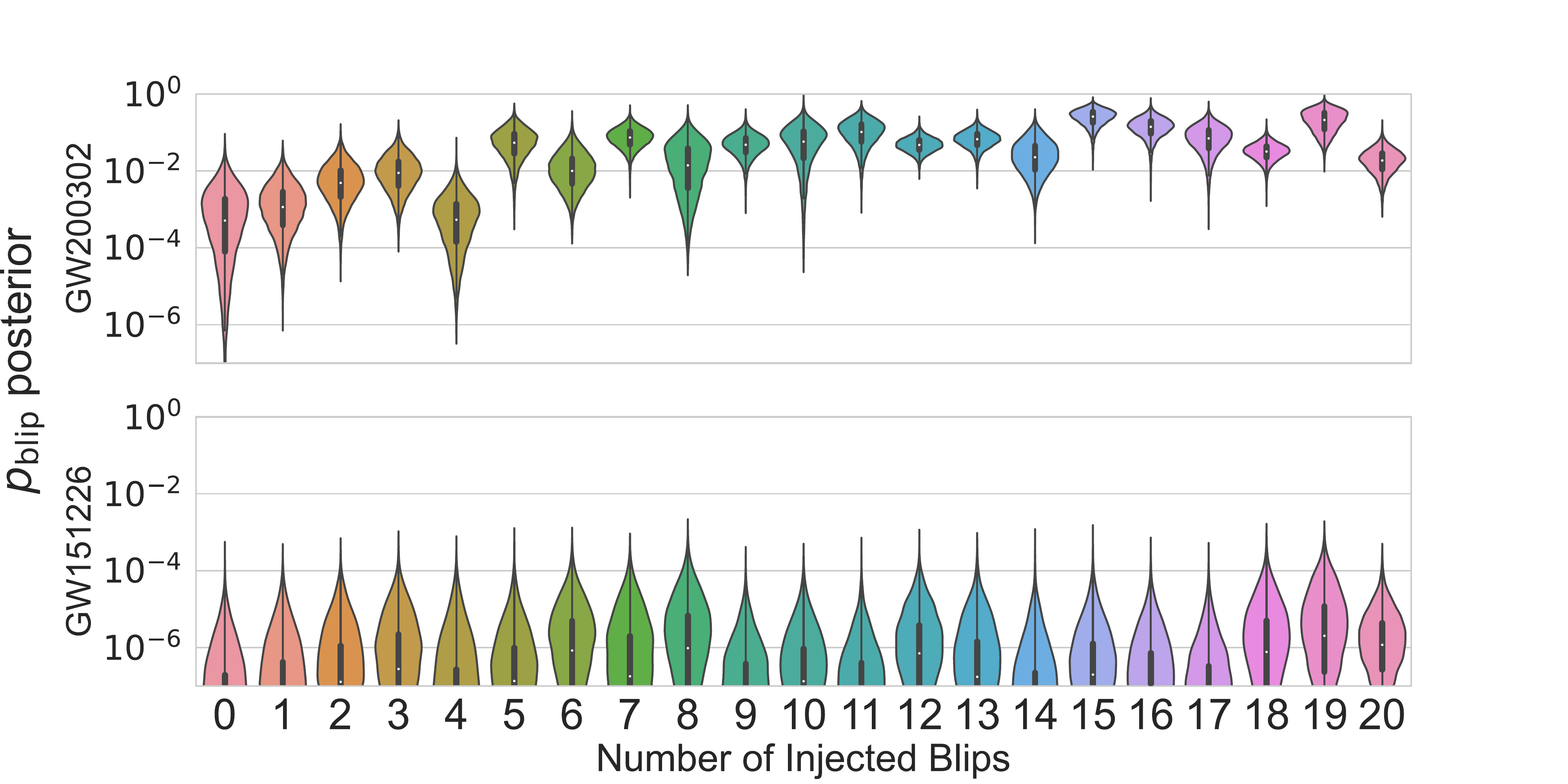}
    \caption{Calculated $1-\pastro = p_{\rm blip}$ for two events, GW200302 and GW151226, in each inference. GW200302 consistently had the lowest $\pastro$ of all the BBH events, while we selected GW151226 to be a representative event for the standard BBH in the catalog. Note a subtle trend for decreasing $\pastro$ as the number of injected blips increased.}
    \label{fig:pastro}
\end{figure}
We show posteriors on $1-\pastro = p_{\rm blip}$ for two example events in Fig. \ref{fig:pastro}, GW151226 and GW200302. GW200302 is the event with the highest probability of being a blip, see the appendix for details. Note as the number of injected blips increases, the $p_{\rm blip}$ increases for GW200302. This is because the $\etabar$ posterior converges on lower mixing fractions; lowering the odds that any given event is astrophysical. This is much more apparent in GW200302, where $p_{\rm blip}$ is mostly dominated by these odds. GW151226 is a representative event for what most BBH $\pastro$ posteriors look like. In fact, many posteriors are even more extreme than GW151226; $\log_{10}(p_{\rm blip}) \to -\infty$ for many events, see Tab. \ref{tab:gw_pastros} in the appendix for the full event list. 

Most $\pastro$ posteriors are sharply peaked, nearly delta functions. Translating this into a calculation on the number of BBHs and blips in the catalog, this suggests that the inferred number of BBHs and blips in the catalog is also sharply peaked. Indeed, using the $\pastroi$ defined in Eq. \ref{eq:pastro} we may calculate the probability that exactly $k$ of $N_{\rm events}$ are astrophysical. Since each data realization is independent, the $\pastroi$ of each event will be statistically independent. The probability that exactly $k$ of $N_{\rm events}$ total events in the catalog are BBHs is then
\beq
p_k(\Lambda) = \sum_{\gamma\in\Gamma(k,N_{\rm events})}\left[ \prod_{j=1}^k \pastroyj(\Lambda)\prod_{j=k+1}^{N_{\rm events}} 1-\pastroyj(\Lambda)\right]
\label{eq:p_k}
\eeq
where $\Gamma(k,N_{\rm events})$ is the set of $k$-combinations of $N_{\rm events}$ (it contains $N_{\rm events}$ choose $k$ elements), a subset of the set of permutations of $N_{\rm events}$. Thinking of permutations as one-to-one and onto functions from the set $\{1, ..., N_{\rm events}\}$ to itself, $k$-combinations are permutations where two permutations $\gamma_1$ and $\gamma_2$ are equivalent if there is the set equality $\gamma_1(\{1,..,k\}) = \gamma_2(\{1,..,k\})$. Informally, the probability that exactly $k$ of $N_{\rm events}$ are BBHs is the probability a specific set of $k$ events are BBHs and the others are glitches, summed over all the possible sets of $k$ events. Note that if all $\pastroi$ are the same, Eq. \ref{eq:p_k} reduces to the binomial distribution as expected. However, Eq. \ref{eq:p_k} is much too computationally expensive to evaluate directly. We use a trick with symmetric polynomials to vastly simplify the calculation, see the appendix for details. We also note that \cite{2020PhRvD.102h3026G} consider the sum of the $\pastroi$. This is the expectation value over $k$ of Eq. \ref{eq:p_k}, which is also discussed in further detail in the appendix.

\begin{figure}
    \centering
    \includegraphics[width=0.9\linewidth]{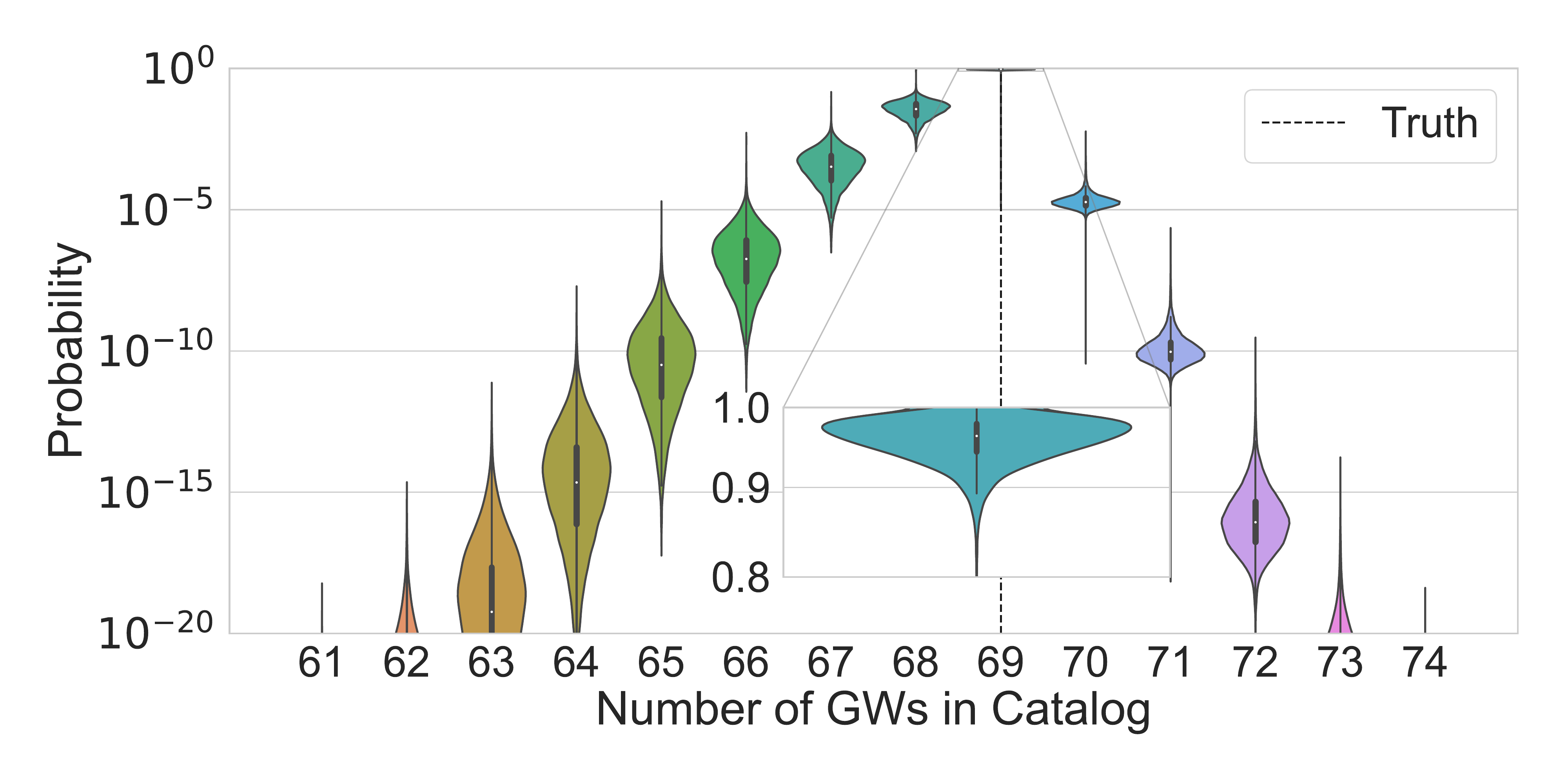}
    \caption{The probability of having $k$ events which are astrophysical in the catalog. The horizontal axis is the number of events in the catalog, and the vertical axis represents the $p_k(\Lambda)$ probability of there being exactly $k$ astrophysical events in the catalog. Since the probability for exactly 69 BBHs rails against 1, we show an inset zoom on the $p_{69}(\Lambda)$ violin. The uncertainty in the value of the probability $p_k(\Lambda)$ comes from the uncertainty in the population parameters $\Lambda$. This particular run was with 69 BBHs injected and 20 contaminant blips injected.}
    \label{fig:p_n_astro}
\end{figure}

\begin{figure}
    \centering
    \includegraphics[width=0.95\linewidth]{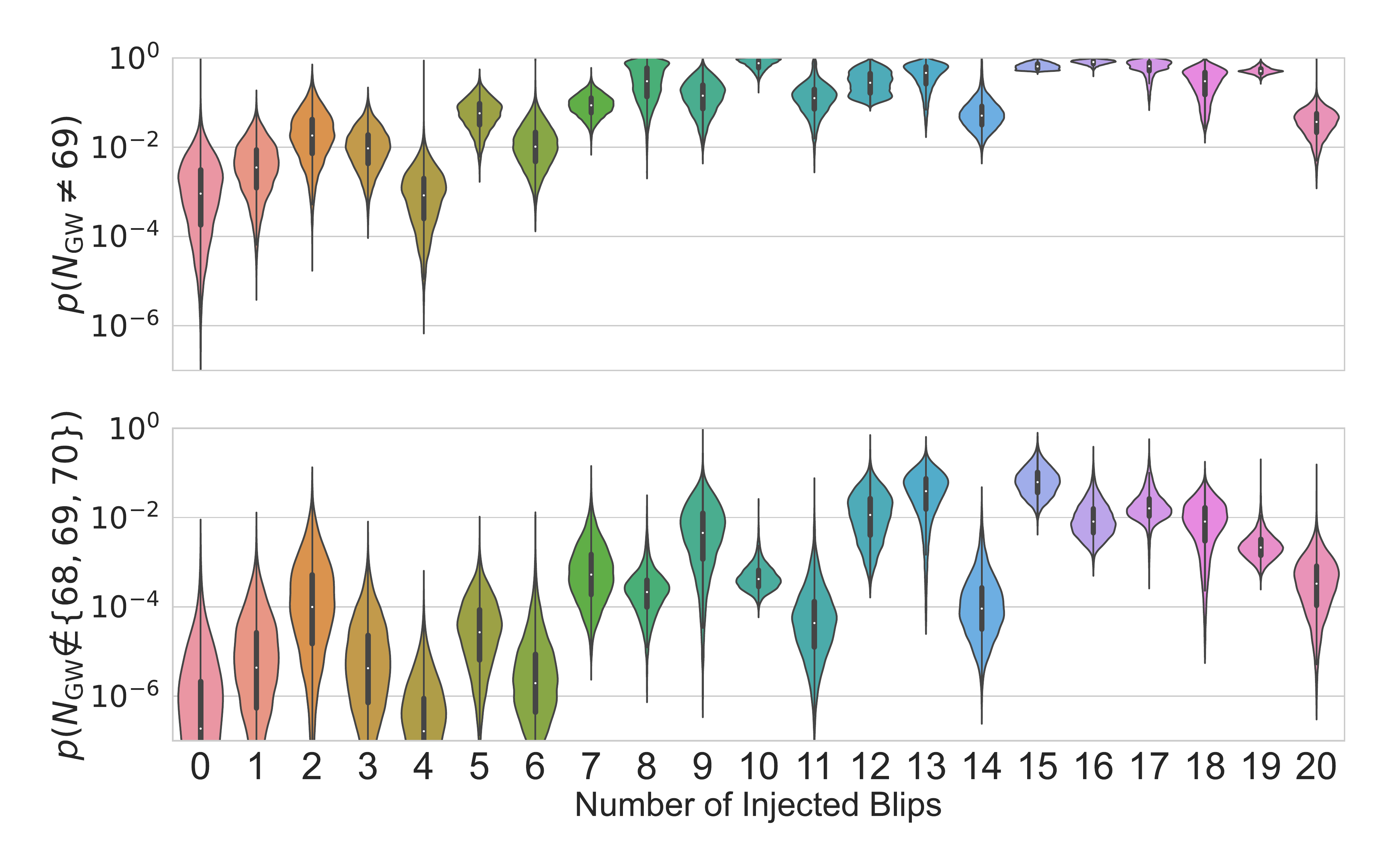}
    \caption{In the top panel, we show violins for the inferred posterior probabilities of the catalog not having 69 BBHs in it; $1-p_{69}(\Lambda)$. The vertical axis shows the logarithm of the probability, and the horizontal axis is the number of injected blips in the catalog. In the bottom panel, we show the posterior probabilities of the catalog having some number of BBHs which is not 68, 69, or 70; $1-p_{68}(\Lambda) - p_{69}(\Lambda) - p_{70}(\Lambda)$. Note the increase in the probabilities as the number of injected blips increases; this is due to higher odds that any given event is a blip (lower $\etabar$). The dip at exactly 20 injected blips is because those 20 contaminants happen to be easily resolvable from the GW population, and so $p_{69}(\Lambda)$ peaks strongly at 1.}
    \label{fig:p_n_astro_all}
\end{figure}

After contaminating the catalog of 69 BBHs passing the LVK selection criteria \citep{PhysRevX.13.011048} with 0 - 20 independently drawn random blips, and running 21 inferences on the hyperparameters $\Lambda$ on the 21 variably-contaminated catalogs, we calculate Eq. \ref{eq:p_k} for each $\Lambda$ sample. We show an example in Fig. \ref{fig:p_n_astro}, the run with 20 contaminant blips. In this run and in most runs, the probability for exactly 69 BBHs in the catalog rails against 1, while for some other runs it can be more uncertain. Variability between runs is due to the differences in how ``BBH-like'' the blip contaminants are, and how well they fit into the blip population model.  
We also show posteriors on the probabilities of having exactly 69 BBHs in the catalog. Specifically, since many of the probabilities rail against 1, we show the logarithm of the negation: the $\log_{10}$ probability of not having 69 BBHs in the catalog, shown in the top panel of Fig. \ref{fig:p_n_astro_all}. As the number of contaminants increases, the resolving power drops, meaning the probability becomes more spread out between $\sim 68-70$. Furthermore, the odds any given event is a BBH drops, as the mixing fraction between BBHs and blips becomes more blip-favored. That said, up to 20 injected blips we observe significant probabilities of exactly 69 BBHs in the catalog, and near unity probabilities of 68 or 69 or 70 BBHs in the catalog (Fig. \ref{fig:p_n_astro_all}). 
While there is some variation in the probabilities, this method consistently recovers the correct number of injected contaminants, so long as the populations are sufficiently dissimilar. It is not clear that the correctly recovering the number of contaminants prevents slight biases from arising in the population inference, especially given there is some small variability in the inferred number of contaminants in the catalog.

\subsection{Biases in the BBH Population}

\begin{figure}
    \centering
    \includegraphics[width=0.95\linewidth]
    {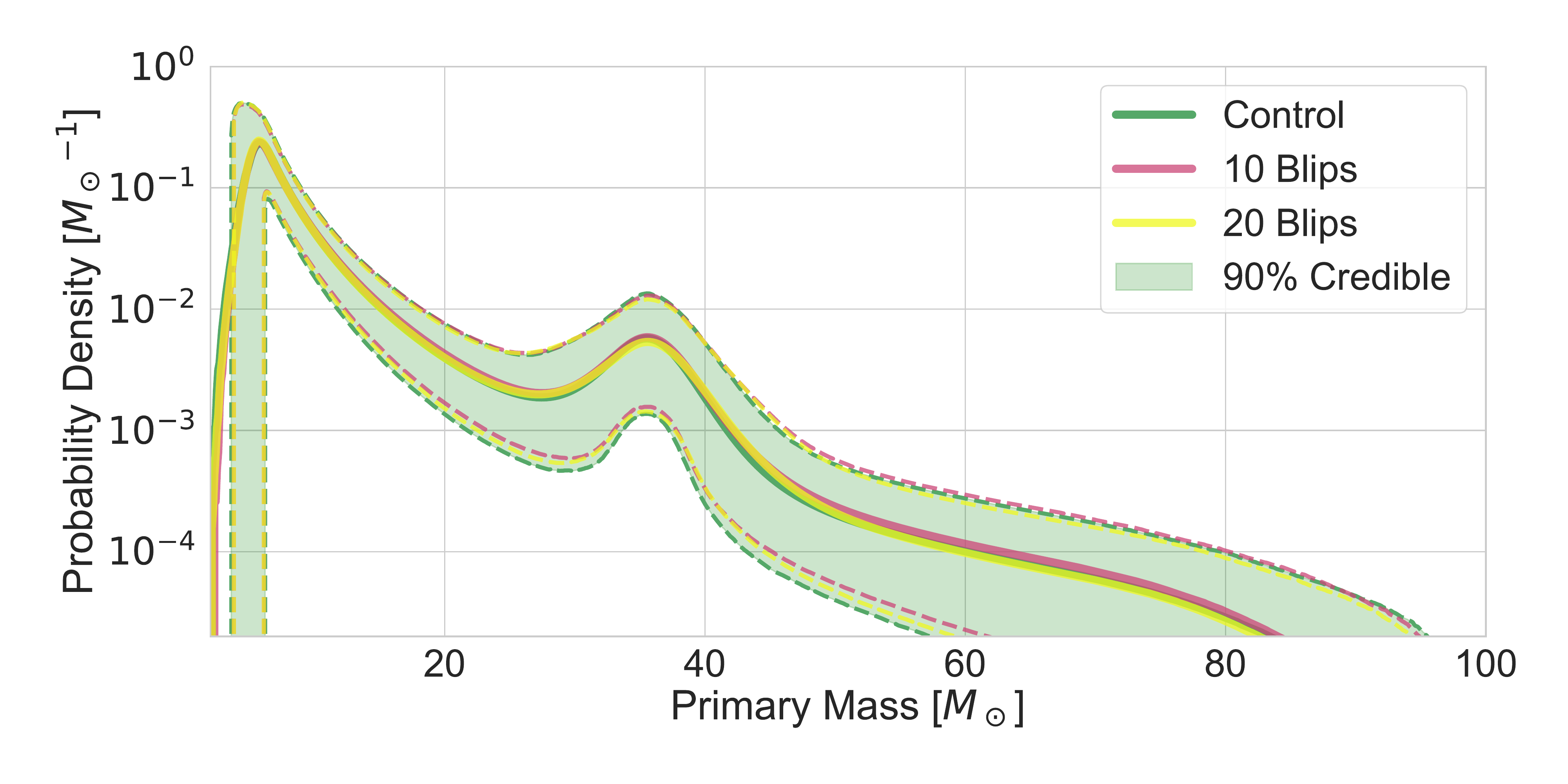}
    \caption{The inferred astrophysical mass distribution. In green we show the control run, with no contaminants injected and no glitch model included. We also show the runs with with the glitch model included and injected contaminants; we show runs with 10 and 20 blips included. The solid line is the posterior population distribution (PPD) and the dashed lines show the upper and lower limits on the 90\% credible region. The inferred distributions appear consistent.}
    \label{fig:inferred_m1s}
\end{figure}

While the correct number of blips is recovered in each run, we want to be sure that no biases are introduced in the inferred astrophysical distributions. For example, we show inferred distributions of the primary masses for a control run and with 10 and 20 injected blips in Fig. \ref{fig:inferred_m1s}. Qualitatively speaking, they appear to be essentially identical. The control run is a population inference on the catalog of 69 BBHs in \cite{PhysRevX.13.011048}, using the same astrophysical population model parameterization described in section \ref{sec:glitchpop}. 
We quantify any differences by calculating the Jensen-Shannon (JS) divergence between the inferred distributions of a control population inference and the inferred astrophysical sub-populations from contaminated catalogs. The JS divergences show no trends, with a median consistently at $\sim 0.09-0.1$ bits. We show the JS divergences in the middle column of Table \ref{tab:js_divergences} in the appendix, and in the first row we show the JS divergences between two draws from the control hyperparameters.

\subsection{Biases from Unmodeled Blip Contaminants}

\begin{figure}
    \centering
    \includegraphics[width=0.98\linewidth]{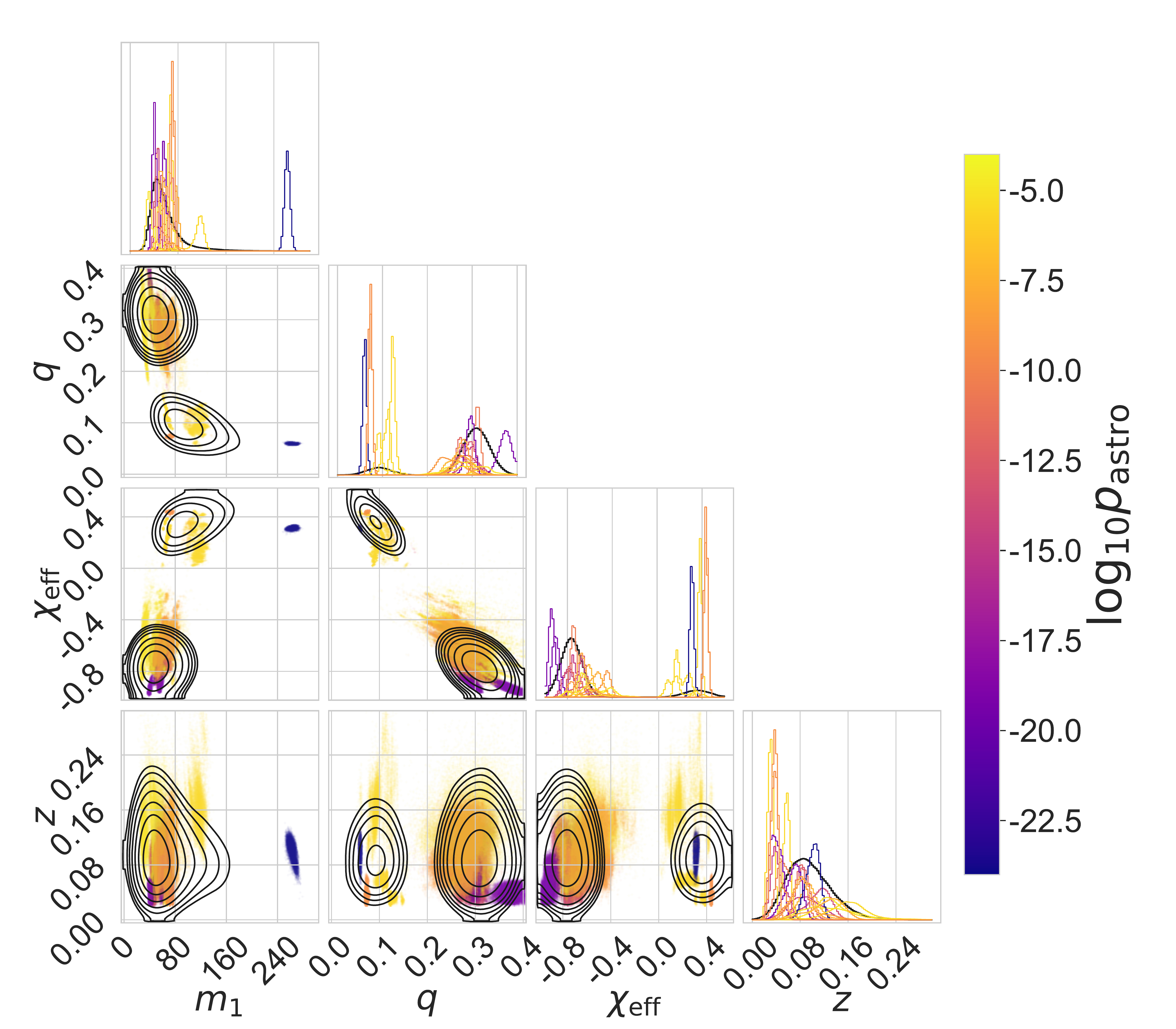}
    \caption{We show the 20 posteriors on the blip effective BBH parameters injected, and their corresponding mean $\pastro$, labelled in the figure by the color of the posterior points. See the colorbar on the right. Note some general patterns: very low $\chieff$ values and very high masses correspond to low $\pastro$ values. Note also that all the blip $\pastro$ values are still very low, less than $10^{-4}$.}
    \label{fig:blip20_ppd_overlay}
\end{figure}

Some glitches appear significantly more astrophysical than others. For the run with 20 blips injected and the 69 BBH mergers, we plot the posteriors on the effective ``BBH'' parameters of the glitches, and population-averaged $\pastro$ values overlaid on the blip PPD, see Fig. \ref{fig:blip20_ppd_overlay}. 
There are some general patterns, most notably that extreme $\chieff$ seems to be the strongest predictor of low $\pastro$, and if the primary mass $m_1$ falls above the maximum mass cutoff $m_{\rm max}$ in the astrophysical model, the $\pastro$ is zero. We show a table of the median and 90\% credible region parameters of each blip, along with the SNR and $\pastro$ in Table \ref{tab:blip_params} in the appendix.

We want to understand the kind of biases which are induced by including blips into the population, without controlling for those contaminants with a glitch model. 
Of the run with 20 injected blips and 69 GWs, we select the blips which could most plausibly be astrophysical, ie. they have the highest $\pastro$. We selected the blip with the highest $\pastro$ (the top row in Table \ref{tab:blip_params}), and the 10 blips with the highest $\pastro$ (the top 10 rows in Table \ref{tab:blip_params}), and contaminated the catalog of 69 BBH mergers passing the LVK selection criteria \citep{PhysRevX.13.011048} with these 1 and 10 blips. We then sample from the population hyperposterior without any glitch model. 

In order to prevent population hyperparameters from railing against prior ranges, we extended the prior range of $m_{\rm max}$ significantly (the maximum cutoff mass parameter in the model of \citealt{2018ApJ...856..173T}) to allow values up to $500 M_{\odot}$.

\begin{figure}
    \centering
    \includegraphics[width=0.95\linewidth]{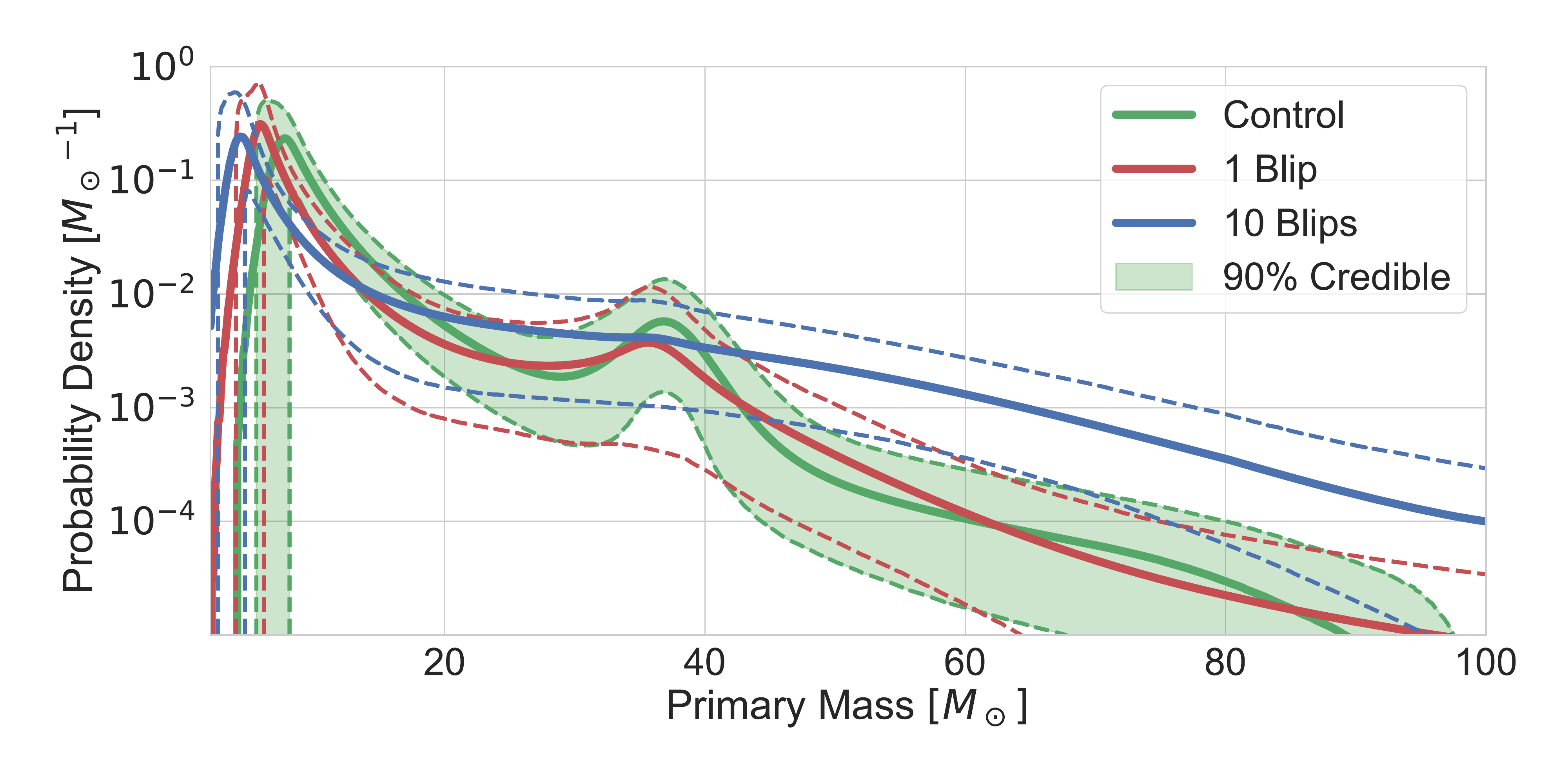}
    \caption{The inferred mass distribution for a control run compared to the inferred mass distribution when 1 and 10 astrophysically plausible blips are included into the catalog, without controlling for their bias with a glitch model. Note the increased support at high mass, and the broadening of the gaussian peak. The low mass end of the distribution is much less affected.}
    \label{fig:bias}
\end{figure}

All the inferred distributions are biased. For instance, we show the inferred primary mass distribution for the control run, and for 1 and 10 contaminants, see Fig. \ref{fig:bias}.
We compute the Jensen-Shannon divergences for these inferred distributions, compared to the control distribution. We show them in the right hand column of Table \ref{tab:js_divergences} in the appendix.

\section{Conclusion and Future Work}
\label{sec:conclusion}
In this article we presented a method for inference of a population of GW sources which is contaminated by non-astrophysical events. We contaminated the catalog of 69 BBHs of \cite{PhysRevX.13.011048} with an increasing number of single-interferometer blip glitches from \cite{2022CQGra..39q5004A}. We showed how to generalize a population inference to not only infer the shape parameters of a GW population, but to simultaneously infer the population of the glitch background events. We tested this method, and showed that it in practice identifies and removes systematic biases from population inference. 
As GW astronomy matures, interesting results may reveal themselves only on the level of populations, and satisfactory statistical significance may require delving into sub-threshold events. 

As a proof of principle analysis, we chose only to consider the blip glitch class from \textsc{GravitySpy}, since \cite{2022CQGra..39q5004A} had already produced parameter estimation samples for these. We caution that the method we presented here will only be robust to blip glitch contamination; we leave it to a future study to do a full simultaneous analysis with a model for an extended population of glitches. 

There is another caveat, in the appropriate estimation of the selection effects. In an end-to-end analysis, the detection criterion is the same for glitches and GWs, and so must be estimated consistently. The current most common method requires a massive set of simulated GWs from a population similar to the population of astrophysical GWs into detector noise, and re-weighting for different population hyperparameters \citep{PhysRevX.13.011048, LVK_injection_set}. The set of glitches comes from regions of parameter space poorly sampled by the injection set, and so to properly estimate the selection effects, one needs an auxiliary suite of injections over the appropriate regions of parameter space. This is a significant computational expense, although it is regularly done by the LVK collaboration to estimate the selection effects of astrophysical GWs.

Though it is a challenge, there are many applications for a method to simultaneously infer the population of astrophysical GWs and non-astrophysical glitches. The most immediate application would be to lower the threshold for including a trigger into the catalog, e.g. select on FAR $<2{\rm yr}^{-1}$, or FAR $<5{\rm yr}^{-1}$. There are real GW events lurking below the FAR $<1{\rm yr}^{-1}$ threshold, and these can aid in constraining the population. This would require an accurate model for the glitches that actually pass the threshold, rather than using our fiducial blip glitch model, and while conceptually similar to this work, the full treatment would also require running end-to-end search pipelines on injections from the glitch population. We leave this to a future study. There are other useful applications as well. Some GWs occur while only a single detector is online \citep{2017CQGra..34o5007C,2020ApJ...897..169N,2022CQGra..39u5012C}. These single detector events often cannot enter a catalog for population inference, and so they cannot be used for constraining the population. Our approach of modelling the intrinsic population of glitches is a step towards the use of single detector triggers in population analyses.

This method can also help characterize triggers found in searches for exotic objects. As an example, BBHs beyond the upper mass gap remain elusive \citep{2021ApJ...909L..23E}. The search sensitivity for these objects is reduced by the presence of short duration glitches much like blip glitches \citep{2019CQGra..36o5010C}, and so a joint analysis of a population of these background glitches and the astrophysical ``beyond-the-gap'' BBHs would measure tighter constraints on their rates. As another example, an analogous procedure is conceivable for continuous wave sources. One may be able to characterize the population of continuous waves (CWs) and the ``glitches'' associated, which are due to monochromatic coherent power between detectors \citep{2021A&A...649A..92C, 2020ApJ...902L..21A,2022PhRvD.106j2008A,2022ApJ...935....1A}. This may benefit a search for CWs or population level characterization of CW sources.

For analyses like the one presented, it is critical to have both an accurate waveform model for glitches and an accurate glitch population model. In this paper, we model glitches with a GW waveform, however, it may be useful to use alternative glitch waveforms. One option is to use non-coherent GW waveforms to model the glitches, where the signal in each interferometer is fit with independent GW waveforms \citep{2010PhRvD..81f2003V}. One can also use non-GW waveform models, such as \textsc{Glitschen} \citep{2021PhRvD.104j2004M} or \textsc{BayesWave} \citep{2015CQGra..32m5012C}. In cases where the glitch waveform model is different from the GW waveform, Eq. \ref{eq:hyper-like-glitch} must be used in its more general form. 
Second, we must have an appropriate model for the glitch population, and using as accurate as possible a model will be crucial. For example, if one continues to use a coherent GW waveform, one could fold in the analysis information about extrinsic parameters, e.g. the fact that the population of glitches is not expected to be isotropic~\citep{2020PhRvD.102j2004P,2022arXiv220400968V,2023PhRvD.107d3016E}. We plan to explore both these avenues in a future work.


\section{Acknowledgements}

The authors wish to thank Sylvia Biscoveanu, Tom Callister, Tom Dent, Reed Essick, Will Farr, and Jacob Golomb for valuable suggestions and insights, and the rates and populations group of the LIGO and
Virgo Collaborations for helpful feedback on this work. The authors thank Michael Zevin and Christopher Berry for comments, edits and feedback.
JH is supported by the National Science Foundation Graduate Research Fellowship Program under Grant No. 2141064. 
SV is partially supported by NSF through the award PHY-2045740.
CT is supported by an MKI Kavli Fellowship.
GA thanks the UKRI Future Leaders Fellowship for support through the grant
MR/T01881X/1. JH and CT gratefully acknowledge the hospitality of Royal Holloway, University of London, where a part of this work was completed. 
This research has made use of data,
software and/or web tools obtained from the Gravitational Wave Open Science Center (https://www.gwopenscience.org), a service of LIGO Laboratory, the
LIGO Scientific Collaboration and the Virgo Collaboration. Virgo is funded by the French Centre National
de Recherche Scientifique (CNRS), the Italian Istituto
Nazionale della Fisica Nucleare (INFN) and the Dutch
Nikhef, with contributions by Polish and Hungarian institutes. We are also grateful to computing resources
provided by the LIGO Laboratory computing clusters
at California Institute of Technology and LIGO Hanford
Observatory supported by National Science Foundation
Grants PHY-0757058 and PHY-0823459. The majority
of analysis performed for this research was done using
resources provided by the Open Science Grid,
which is supported by the National Science Foundation
award \#2030508. 
This material is based upon work supported by NSF’s LIGO Laboratory which is a major facility fully funded by the National Science Foundation.

\section{Data Availability}

The data underlying this work can be found on a zenodo data release at https://doi.org/10.5281/zenodo.7860652. The public gravitational wave data can be found in the Gravitational Wave Open Science Center \citep{theligoscientificcollaboration2023open}.

\bibliographystyle{mnras}
\bibliography{references}

\section*{Appendix}

\subsection*{Event Information}

We show upper bounds on the calculated $\pastro$ for each GW event when we included 20 contaminant blips in Table \ref{tab:gw_pastros}. We also show upper bounds on each $\pastro$, given by the $90\%$ and $99\%$ upper bounds. In Table \ref{tab:blip_params} we show the parameters of the 20 blips that contaminate the catalog in the run with 20 blips. Note we use a random set of blips for each catalog, e.g. the 19 contaminants for the run with 19 blips are not a subset of the 20 contaminants for the run with 20 blips.  
\renewcommand{\arraystretch}{1.5}
\begin{table*}
    \centering
    \begin{tabular}{|c|c|c||c|c|c|}
\hline Event & 90\% Upper bound & 99\% Upper bound & Event & 90\% Upper bound & 99\% Upper bound \\ 
\hline GW150914 & $0$ & $0$ & GW190731\_140936 & $6.7\times 10^{-16}$ & $2.3\times 10^{-13}$ \\ 
\hline GW151012 & $3.1\times 10^{-4}$ & $6.1\times 10^{-4}$ & GW190803\_022701 & $2.3\times 10^{-10}$ & $6.8\times 10^{-9}$ \\ 
\hline GW151226 & $1.2\times 10^{-5}$ & $5.9\times 10^{-5}$ & GW190805\_211137 & $0$ & $0$ \\ 
\hline GW170104 & $1.3\times 10^{-6}$ & $4.8\times 10^{-6}$ & GW190828\_063405 & $0$ & $0$ \\ 
\hline GW170608 & $7.3\times 10^{-8}$ & $4.4\times 10^{-7}$ & GW190828\_065509 & $3.7\times 10^{-9}$ & $3.4\times 10^{-8}$ \\ 
\hline GW151226 & $9.7\times 10^{-4}$ & $2.9\times 10^{-3}$ & GW190910\_112807 & $0$ & $0$ \\ 
\hline GW170809 & $2.6\times 10^{-10}$ & $4.4\times 10^{-9}$ & GW190915\_235702 & $2.3\times 10^{-11}$ & $3.6\times 10^{-10}$ \\ 
\hline GW170814 & $0$ & $0$ & GW190924\_021846 & $8.7\times 10^{-8}$ & $1.1\times 10^{-6}$ \\ 
\hline GW170818 & $0$ & $5.1\times 10^{-15}$ & GW190925\_232845 & $1.7\times 10^{-7}$ & $8.6\times 10^{-7}$ \\ 
\hline GW170823 & $1.6\times 10^{-7}$ & $9.0\times 10^{-7}$ & GW190929\_012149 & $1.3\times 10^{-7}$ & $1.6\times 10^{-6}$ \\ 
\hline GW190408\_181802 & $0$ & $0$ & GW190930\_133541 & $8.2\times 10^{-11}$ & $8.4\times 10^{-10}$ \\ 
\hline GW190412 & $7.2\times 10^{-5}$ & $2.8\times 10^{-4}$ & GW191103\_012549 & $1.8\times 10^{-8}$ & $1.6\times 10^{-7}$ \\ 
\hline GW190413\_052954 & $1.8\times 10^{-9}$ & $3.3\times 10^{-8}$ & GW191105\_143521 & $4.2\times 10^{-9}$ & $3.2\times 10^{-8}$ \\ 
\hline GW190413\_134308 & $3.2\times 10^{-14}$ & $1.7\times 10^{-12}$ & GW191109\_010717 & $6.0\times 10^{-5}$ & $4.6\times 10^{-4}$ \\ 
\hline GW190421\_213856 & $1.9\times 10^{-10}$ & $4.0\times 10^{-9}$ & GW191127\_050227 & $1.8\times 10^{-4}$ & $7.7\times 10^{-4}$ \\ 
\hline GW190503\_185404 & $4.0\times 10^{-2}$ & $7.0\times 10^{-2}$ & GW191129\_134029 & $1.6\times 10^{-11}$ & $2.3\times 10^{-10}$ \\ 
\hline GW190512\_180714 & $2.0\times 10^{-15}$ & $1.4\times 10^{-13}$ & GW191204\_171526 & $0$ & $0$ \\ 
\hline GW190513\_205428 & $4.0\times 10^{-13}$ & $2.8\times 10^{-11}$ & GW191215\_223052 & $0$ & $2.2\times 10^{-16}$ \\ 
\hline GW190517\_055101 & $1.1\times 10^{-8}$ & $3.6\times 10^{-7}$ & GW191216\_213338 & $3.2\times 10^{-6}$ & $1.8\times 10^{-5}$ \\ 
\hline GW190519\_153544 & $0$ & $0$ & GW191222\_033537 & $3.6\times 10^{-11}$ & $1.1\times 10^{-9}$ \\ 
\hline GW190521 & $5.6\times 10^{-16}$ & $4.9\times 10^{-14}$ & GW191230\_180458 & $3.4\times 10^{-7}$ & $4.3\times 10^{-6}$ \\ 
\hline GW190521\_074359 & $0$ & $0$ & GW200112\_155838 & $0$ & $0$ \\ 
\hline GW190527\_092055 & $1.8\times 10^{-9}$ & $2.1\times 10^{-8}$ & GW200128\_022011 & $2.2\times 10^{-16}$ & $2.0\times 10^{-14}$ \\ 
\hline GW190602\_175927 & $1.8\times 10^{-10}$ & $1.2\times 10^{-8}$ & GW200129\_065458 & $4.8\times 10^{-9}$ & $4.7\times 10^{-8}$ \\ 
\hline GW190620\_030421 & $2.4\times 10^{-12}$ & $1.8\times 10^{-10}$ & GW200202\_154313 & $1.9\times 10^{-7}$ & $1.1\times 10^{-6}$ \\ 
\hline GW190630\_185205 & $0$ & $0$ & GW200208\_130117 & $2.0\times 10^{-7}$ & $1.7\times 10^{-6}$ \\ 
\hline GW190701\_203306 & $4.8\times 10^{-12}$ & $1.6\times 10^{-10}$ & GW200209\_085452 & $7.7\times 10^{-9}$ & $1.1\times 10^{-7}$ \\ 
\hline GW190706\_222641 & $2.3\times 10^{-15}$ & $5.4\times 10^{-13}$ & GW200216\_220804 & $4.1\times 10^{-7}$ & $2.3\times 10^{-6}$ \\ 
\hline GW190707\_093326 & $4.6\times 10^{-11}$ & $4.8\times 10^{-10}$ & GW200219\_094415 & $1.3\times 10^{-12}$ & $1.1\times 10^{-10}$ \\ 
\hline GW190708\_232457 & $0$ & $6.7\times 10^{-16}$ & GW200224\_222234 & $1.2\times 10^{-6}$ & $5.4\times 10^{-6}$ \\ 
\hline GW190719\_215514 & $7.5\times 10^{-12}$ & $2.3\times 10^{-10}$ & GW200225\_060421 & $4.7\times 10^{-8}$ & $3.7\times 10^{-7}$ \\ 
\hline GW190720\_000836 & $1.4\times 10^{-10}$ & $1.4\times 10^{-9}$ & GW200302\_015811 & $4.1\times 10^{-2}$ & $7.2\times 10^{-2}$ \\ 
\hline GW190725\_174728 & $2.3\times 10^{-6}$ & $8.5\times 10^{-6}$ & GW200311\_115853 & $4.9\times 10^{-4}$ & $1.4\times 10^{-3}$ \\ 
\hline GW190727\_060333 & $0$ & $0$ & GW200316\_215756 & $1.6\times 10^{-8}$ & $1.2\times 10^{-7}$ \\ 
\hline GW190728\_064510 & $4.8\times 10^{-12}$ & $7.4\times 10^{-11}$ &  &  &  \\ \hline
    \end{tabular}
    \caption{Inferred $1-\pastro = p_{\rm blip}$ for each event in the catalog of \protect\cite{PhysRevX.13.011048}, calculated from the run with 20 injected blips. We show the upper bounds on the inferred $p_{\rm blip}$ at both 90\% and 99\% credence. Note GW200302 has $p_{\rm blip} \lesssim 7.2\%$, the highest non-astrophysical probability event, and the event GW190503\_185404 has the second highest $p_{\rm blip} \lesssim 7.0\%$.}
    \label{tab:gw_pastros}
\end{table*}

\subsection*{Optimal Detectable Mixing Fraction Posterior}

Consider the scenario where the populations are disjoint such that every event posterior uniquely determines which population the event originates from. The event parameters tell us with no ambiguity whether an event is a glitch or a GW. Therefore, we want to infer the relative rate of detectable events given we detected $N_{\rm events}$, with $k$ unambiguous astrophysical events, and the rest unambiguous glitches.
This is a common problem in Bayesian inference and it admits an analytical posterior, given by Eq. \ref{eq:beta_model}. This is the best the inference could possibly constrain $\etabar$, and so it is a useful benchmark to compare to. 
\beq
p(\etabar) = \frac{\etabar^k(1-\etabar)^{N_{\rm events} - k}}{B(k+1,N_{\rm events} -k + 1)},
\label{eq:beta_model}
\eeq
which assumes a uniform prior in $\etabar$ from 0 to 1, and the denominator is a normalization.

In fact, we can see how this arises directly from Eq. \ref{eq:hyper-like-glitch-alt}. If the population models for the glitches and the BBHs are completely disjoint for all event posteriors, then in each term in the product of Eq. \ref{eq:hyper-like-glitch-alt}, either the glitch term $\int d\psi \mathcal{L}(d_i|\psi)p_G(\psi|\Lambda_G)$ or the astrophysical term $\int d\theta \mathcal{L}(d_i|\theta)p_A(\theta|\Lambda_A)$ will vanish. The likelihood then factorizes:
\begin{align}
&\mathcal{L}(\{d_i\} | \Lambda_A, \Lambda_G, \eta) \propto  \etabar^k (1-\etabar)^{N_{\rm events} - k} \nn \\ &\prod_{i=1}^k \frac{\int d\theta \mathcal{L}(d_i|\theta)p_A(\theta|\Lambda_A)}{\alpha_A(\Lambda_A)}\prod_{i=k+1}^{N_{\rm events}}\frac{\int d\psi \mathcal{L}(d_i|\psi)p_G(\psi|\Lambda_G)}{\alpha_G(\Lambda_G)},
\label{eq:hyper-like-glitch-separate}    
\end{align}
and so the inference may proceed independently for the astrophysical hyperparameters $\Lambda_A$, the glitch hyperparameters $\Lambda_G$ and the detectable mixing fraction $\etabar$. This matches the intuitive result that independent populations may be characterized independently. Pulling out the $\etabar$ term in the likelihood and normalizing with a uniform prior between 0 and 1, we recover Eq. \ref{eq:beta_model}.

In general, the glitch and BBH populations are not completely disjoint and the glitch/astrophysical terms in the product in Eq. \ref{eq:hyper-like-glitch-alt} do not vanish. With the additional uncertainty in the ``identity'' of each event in the catalog, the posterior on $\etabar$ will broaden. The degree of broadening tells us how close the inferences are coming towards knowing there are exactly $k$ BBHs of $N_{\rm events}$ total events.

\subsection*{GW200302}

\begin{figure}
    \centering
    \includegraphics[width=0.98\linewidth]{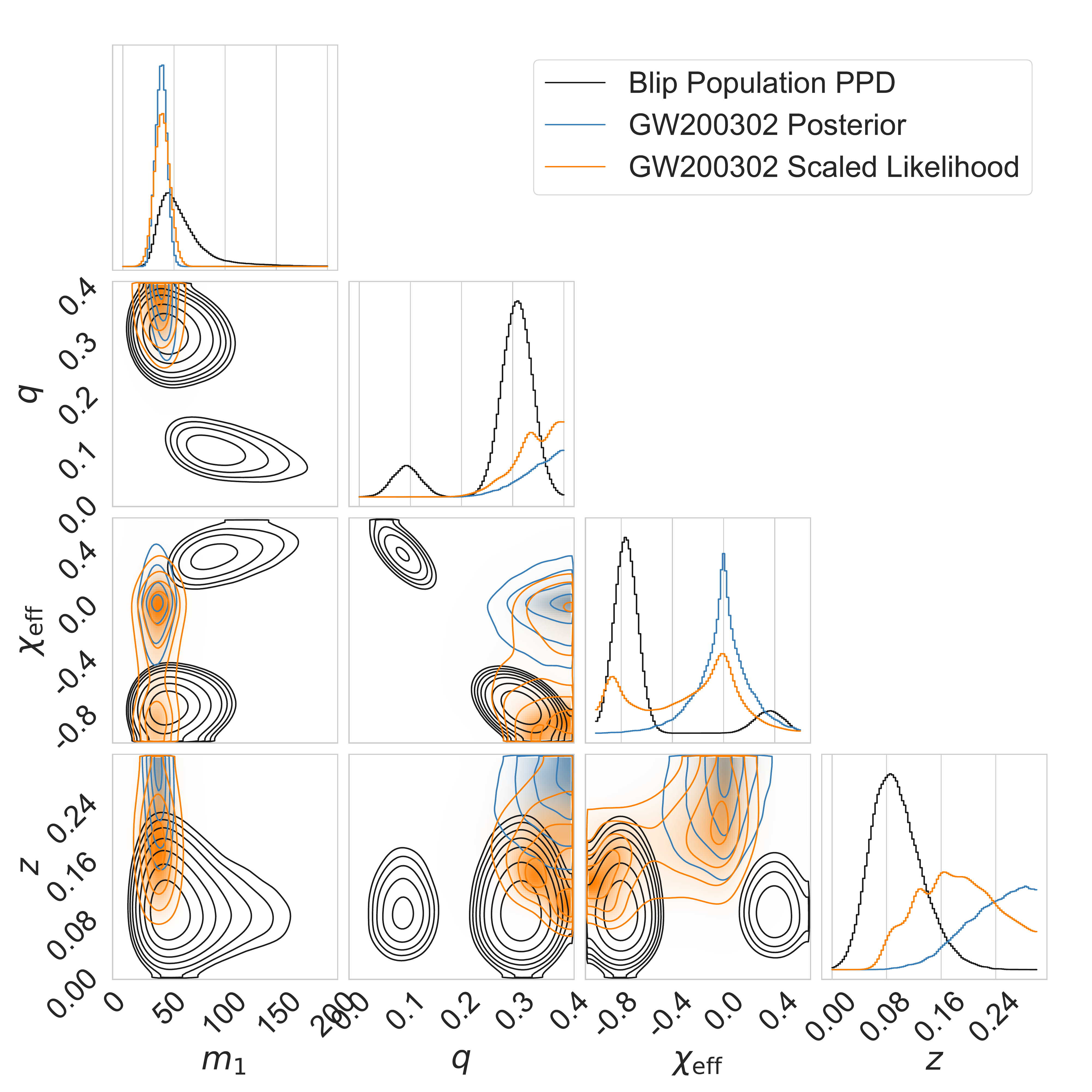}
    \caption{A corner plot with the primary mass, mass ratio, effective aligned spin, and redshift posterior of GW200302 overlaid on the population predictive distribution of the blip population. The posterior on GW200302 is shown in blue, with the first 4$\sigma$ contours and sample points. The blip population predictive distribution is in black contours, showing the first 7$\sigma$. Since it is not the posterior overlap but the likelihood overlap which contributes to the $p_{\rm blip}$, we include the posterior reweighted by the inverse of the prior. This highlights the regions of high overlap for the glitch likelihood term.}
    \label{fig:GW200302}
\end{figure}

GW200302 has the largest support for $1 - \pastro = p_{\rm blip}$. To understand this, we show the corner plot overlay of the GW200302 posterior and the blip population predictive distribution in Fig. \ref{fig:GW200302}. Note the tails of the GW200302 posterior overlaps with the blip population distribution; this is why the $\pastro$ for GW200302 is relatively low.

Further, it is not the posterior ``overlap'' which is taken into the population likelihood, but the event likelihood ``overlap.'' The tails of the posterior in $\chieff$ and in mass ratio $q$--the samples which happen to fall neatly into the blip population--are therefore weighted much more highly, since the sampling prior there is much lower. Indeed the $\chieff$ posterior is essentially the recovered $\chieff$ sampling prior. This drives up the glitch population term in Eq. \ref{eq:pastro}, and therefore lowers the $\pastro$. This is expected: if there are poor constraints on the source parameters, we must be more agnostic about the event's origin based on the parameters alone.

\renewcommand{\arraystretch}{1.5}
\begin{table*}
    \centering
    \begin{tabular}{|c|c|c|c|c|c|c|c|c|}
    \hline
Number & $m_1$ & $q$ & $\chieff$ & $z$ & SNR & Median $\pastro$ & $\pastro$ 5\% & $\pastro$ 95\% 
 \\ \hline  1 & $49.9^{+12.0}_{-5.5}$ & $0.28^{+0.05}_{-0.05}$ & $-0.64^{+0.28}_{-0.09}$ & $0.12^{+0.07}_{-0.06}$ & $10.9^{+1.7}_{-1.8}$ & $1.0\times 10^{-5}$ & $5.2\times 10^{-6}$ & $2.0\times 10^{-5}$ 
 \\ \hline 2 & $31.2^{+12.1}_{-4.1}$ & $0.3^{+0.1}_{-0.1}$ & $-0.58^{+0.23}_{-0.22}$ & $0.13^{+0.06}_{-0.04}$ & $8.62^{+1.83}_{-1.89}$ & $5.8\times 10^{-6}$ & $2.7\times 10^{-6}$ & $1.4\times 10^{-5}$ 
 \\ \hline 3 & $65.1^{+5.6}_{-19.3}$ & $0.11^{+0.2}_{-0.02}$ & $0.19^{+0.12}_{-0.95}$ & $0.058^{+0.01}_{-0.008}$ & $15.4^{+1.7}_{-1.7}$ & $8.3\times 10^{-7}$ & $3.9\times 10^{-8}$ & $1.1\times 10^{-5}$ 
 \\ \hline 4 & $66.6^{+0.8}_{-2.0}$ & $0.12^{+0.01}_{-0.0}$ & $0.38^{+0.0}_{-0.02}$ & $0.031^{+0.007}_{-0.006}$ & $27.5^{+1.6}_{-1.7}$ & $4.2\times 10^{-8}$ & $3.1\times 10^{-11}$ & $5.9\times 10^{-6}$ 
 \\ \hline 5 & $56.4^{+4.8}_{-3.5}$ & $0.26^{+0.03}_{-0.04}$ & $-0.62^{+0.13}_{-0.07}$ & $0.093^{+0.033}_{-0.029}$ & $13.0^{+1.7}_{-1.7}$ & $1.3\times 10^{-8}$ & $1.0\times 10^{-10}$ & $2.7\times 10^{-7}$ 
 \\ \hline 6 & $116.0^{+9.8}_{-40.5}$ & $0.094^{+0.175}_{-0.013}$ & $0.15^{+0.21}_{-0.73}$ & $0.16^{+0.07}_{-0.05}$ & $10.5^{+1.7}_{-1.8}$ & $5.6\times 10^{-9}$ & $3.5\times 10^{-13}$ & $1.1\times 10^{-5}$ 
 \\ \hline 7 & $51.4^{+21.0}_{-2.9}$ & $0.28^{+0.02}_{-0.03}$ & $-0.67^{+0.17}_{-0.06}$ & $0.082^{+0.052}_{-0.031}$ & $14.5^{+1.7}_{-1.8}$ & $2.9\times 10^{-9}$ & $4.8\times 10^{-15}$ & $2.6\times 10^{-8}$ 
 \\ \hline 8 & $76.8^{+5.2}_{-4.7}$ & $0.28^{+0.04}_{-0.04}$ & $-0.54^{+0.11}_{-0.12}$ & $0.13^{+0.04}_{-0.05}$ & $15.5^{+1.7}_{-1.7}$ & $1.1\times 10^{-10}$ & $1.8\times 10^{-18}$ & $5.2\times 10^{-8}$ 
 \\ \hline 9 & $45.9^{+4.4}_{-1.7}$ & $0.31^{+0.01}_{-0.01}$ & $-0.75^{+0.03}_{-0.04}$ & $0.051^{+0.016}_{-0.017}$ & $21.2^{+1.6}_{-1.7}$ & $3.0\times 10^{-12}$ & $2.7\times 10^{-25}$ & $1.2\times 10^{-10}$ 
 \\ \hline 10 & $61.1^{+4.8}_{-7.4}$ & $0.25^{+0.05}_{-0.04}$ & $-0.51^{+0.11}_{-0.28}$ & $0.083^{+0.025}_{-0.029}$ & $14.8^{+1.7}_{-1.7}$ & $3.4\times 10^{-13}$ & $1.1\times 10^{-20}$ & $6.9\times 10^{-9}$ 
 \\ \hline 11 & $69.4^{+5.3}_{-7.0}$ & $0.27^{+0.03}_{-0.03}$ & $-0.72^{+0.16}_{-0.08}$ & $0.079^{+0.016}_{-0.03}$ & $20.6^{+1.7}_{-1.6}$ & $9.7\times 10^{-16}$ & $2.5\times 10^{-23}$ & $2.5\times 10^{-12}$ 
 \\ \hline 12 & $71.9^{+2.9}_{-1.4}$ & $0.074^{+0.003}_{-0.003}$ & $0.44^{+0.01}_{-0.01}$ & $0.037^{+0.005}_{-0.008}$ & $36.0^{+1.7}_{-1.7}$ & $2.0\times 10^{-16}$ & $0$ & $1.0\times 10^{-10}$ 
 \\ \hline 13 & $69.8^{+1.2}_{-1.3}$ & $0.073^{+0.001}_{-0.002}$ & $0.44^{+0.0}_{-0.0}$ & $0.038^{+0.017}_{-0.011}$ & $26.7^{+1.6}_{-1.7}$ & $2.9\times 10^{-17}$ & $0$ & $2.9\times 10^{-11}$ 
 \\ \hline 14 & $43.7^{+11.8}_{-2.1}$ & $0.3^{+0.0}_{-0.0}$ & $-0.67^{+0.06}_{-0.11}$ & $0.084^{+0.019}_{-0.019}$ & $15.1^{+1.7}_{-1.7}$ & $8.3\times 10^{-21}$ & $8.0\times 10^{-32}$ & $3.7\times 10^{-15}$ 
 \\ \hline 15 & $57.3^{+6.1}_{-4.4}$ & $0.28^{+0.04}_{-0.04}$ & $-0.77^{+0.15}_{-0.08}$ & $0.11^{+0.03}_{-0.04}$ & $14.3^{+1.7}_{-1.7}$ & $5.2\times 10^{-21}$ & $6.3\times 10^{-32}$ & $2.2\times 10^{-15}$ 
 \\ \hline 16 & $52.9^{+11.2}_{-6.8}$ & $0.29^{+0.03}_{-0.03}$ & $-0.77^{+0.13}_{-0.09}$ & $0.088^{+0.045}_{-0.041}$ & $14.2^{+1.7}_{-1.7}$ & $7.2\times 10^{-22}$ & $4.0\times 10^{-33}$ & $4.7\times 10^{-16}$ 
 \\ \hline 17 & $45.1^{+2.4}_{-1.9}$ & $0.29^{+0.02}_{-0.02}$ & $-0.71^{+0.07}_{-0.06}$ & $0.041^{+0.023}_{-0.013}$ & $17.9^{+1.7}_{-1.7}$ & $3.0\times 10^{-26}$ & $2.2\times 10^{-42}$ & $5.3\times 10^{-18}$ 
 \\ \hline 18 & $56.3^{+2.6}_{-2.3}$ & $0.3^{+0.0}_{-0.0}$ & $-0.9^{+0.1}_{-0.0}$ & $0.07^{+0.02}_{-0.03}$ & $19.3^{+1.7}_{-1.7}$ & $2.2\times 10^{-36}$ & $6.8\times 10^{-60}$ & $9.0\times 10^{-25}$ 
 \\ \hline 19 & $40.3^{+1.7}_{-1.5}$ & $0.37^{+0.02}_{-0.03}$ & $-0.94^{+0.05}_{-0.03}$ & $0.039^{+0.014}_{-0.012}$ & $22.9^{+1.6}_{-1.7}$ & $4.1\times 10^{-45}$ & $2.0\times 10^{-77}$ & $7.3\times 10^{-29}$ 
 \\ \hline 20 & $262.0^{+5.7}_{-5.5}$ & $0.06^{+0.0}_{-0.0}$ & $0.31^{+0.01}_{-0.01}$ & $0.1^{+0.0}_{-0.0}$ & $19.9^{+1.7}_{-1.7}$ & $0$ & $0$ & $0$ 
 \\ \hline
    \end{tabular}
    \caption{Median and 90\% credible intervals for the GW effective parameters of the 20 blip contaminants for the 20 injection run, organized by the median $\pastro$. The SNR is the optimal SNR.}
    \label{tab:blip_params}
\end{table*}

\subsection*{Jensen Shannon Divergences}

We show the Jensen Shannon divergences measured between the inferred astrophysical sub-populations of each run, where we included the glitch model to account for the injected contaminants. We also show the Jensen Shannon Divergence for 1 and 10 injected blips where we did not include the glitch model. \renewcommand{\arraystretch}{1.5}
\begin{table}
    \centering
    \begin{tabular}{|c|c|c|}
    \hline
        $N_{\rm blip}$ & JS w/ glitch model (bits) & JS w/o glitch model (bits) \\
        \hline Control & $0.097^{+0.123}_{-0.063}$ & $0.097^{+0.123}_{-0.063}$ \\ \hline 
 	0 & $0.088^{+0.119}_{-0.057}$ & $---$ \\ \hline 
 	1 & $0.091^{+0.117}_{-0.058}$ & $0.104^{+0.133}_{-0.065}$ \\ \hline 
 	2 & $0.090^{+0.119}_{-0.059}$ & $---$ \\ \hline 
 	3 & $0.090^{+0.117}_{-0.059}$ & $---$ \\ \hline 
 	4 & $0.088^{+0.118}_{-0.057}$ & $---$ \\ \hline 
 	5 & $0.093^{+0.119}_{-0.059}$ & $---$ \\ \hline 
 	6 & $0.092^{+0.121}_{-0.060}$ & $---$ \\ \hline 
 	7 & $0.091^{+0.126}_{-0.059}$ & $---$ \\ \hline 
 	8 & $0.094^{+0.126}_{-0.059}$ & $---$ \\ \hline 
 	9 & $0.090^{+0.116}_{-0.057}$ & $---$ \\ \hline 
 	10 & $0.093^{+0.118}_{-0.059}$ & $0.406^{+0.188}_{-0.132}$ \\ \hline 
 	11 & $0.088^{+0.113}_{-0.056}$ & $---$ \\ \hline 
 	12 & $0.090^{+0.119}_{-0.057}$ & $---$ \\ \hline 
 	13 & $0.088^{+0.120}_{-0.056}$ & $---$ \\ \hline 
 	14 & $0.090^{+0.120}_{-0.058}$ & $---$ \\ \hline 
 	15 & $0.091^{+0.117}_{-0.059}$ & $---$ \\ \hline 
 	16 & $0.095^{+0.121}_{-0.060}$ & $---$ \\ \hline 
 	17 & $0.097^{+0.120}_{-0.062}$ & $---$ \\ \hline 
 	18 & $0.091^{+0.126}_{-0.057}$ & $---$ \\ \hline 
 	19 & $0.089^{+0.115}_{-0.056}$ & $---$ \\ \hline 
 	20 & $0.090^{+0.119}_{-0.057}$ & $---$ \\ \hline 
    \end{tabular}
    \caption{Jensen-Shannon (JS) divergences in units of bits (base-2 logarithm) between the inferred distributions of a control run and the astrophysical sub-population of the simultaneous fitting runs. In the middle column are the runs which include a glitch model, but in the right column we show runs which do not have a glitch model and so contaminants must be fitted with the astrophysical population. In the control row, we show the JS divergence posterior from two random samples from the hyperparameter posterior in the control run. Note the consistency between each run in the middle column, in particular, the lack of any sort of (increasing) trend. In the right column notice that the JS divergences increase as expected.}
    \label{tab:js_divergences}
\end{table}

\subsection*{Calculating the Inferred Number of Events in the Catalog}

In Eq. \ref{eq:p_k} we show an expression for the probability on the number of events in the catalog, depending on the population hyperparameters $\Lambda$. However, this expression is a sum of $\mathcal{O}(10^{20})$ terms, and as such is not computationally feasible to evaluate. Fortunately, there is a much more efficient method to complete the sum.

The sum has a largest term, which we can easily find by first ordering the list of $\pastroi$ from largest to smallest. This term then corresponds to the identity $k$-combination, denoted $\gamma_0 = I$. We notate this term by $p^{(0)} = \prod_{i=1}^k \pastroi\prod_{i=k+1}^{N_{\rm events}}(1-\pastroi)$. Because many of the $\pastro$ posteriors have non-negligible posterior width, this term does not completely dominate the entire sum, however, we can express all the other $k$-combinations in the sum in terms of this $p^{(0)}$. In particular, we can think of each $\gamma \in \Gamma(k, N_{\rm events})$ in terms of the number $r$ of events which must be exchanged from the astrophysical bin to the glitch bin in order to match $\gamma_0$. Because we are summing unique $k$-combinations, the probability associated with the family of $k$-combinations which are $r$ events different from $\gamma_0$ is given by
\begin{align}
    p^{(r)} = p^{(0)}&e_r\left(\frac{1-p_{{\rm astro},1}}{p_{{\rm astro},1}}, ..., \frac{1-p_{{\rm astro},k}}{p_{{\rm astro},k}}\right) \nn \\ &\times e_r\left(\frac{p_{{\rm astro},k+1}}{1-p_{{\rm astro},k+1}}, ..., \frac{p_{{\rm astro},N_{\rm events}}}{1-p_{{\rm astro}, N_{\rm events}}}\right)
\end{align}
where the $e_r$ is the $r^{\rm th}$ symmetric polynomial. Symmetric polynomials are defined such that every term has degree $r$ and every $r$ combination of the variables appears once in the sum, e.g. $e_2(x,y,z) = xy + xz + yz$. Note how symmetric polynomials naturally capture the idea of summing over unique sets. The first polynomial term is the sum over all unique sets of size $r$ sending events from the astrophysical bin to the glitch bin. The second polynomial term is similar, sending all unique sets of size $r$ from the glitch bin to the astrophysical bin. Their product, then, is the sum over all combinations of unique set exchanges of size $r$ between the astrophysical and glitch bin. 

This is nice, but it is not helpful unless one can rapidly evaluate the symmetric polynomials. It turns out that one can easily find the $r^{\rm th}$ symmetric polynomial recursively from the previous $r-1$ symmetric polynomials, using Newton and Girard's Theorem:
\beq
r e_r(x_1, ..., x_n) = \sum_{j=1}^r (-1)^{j-1} e_{r-j}(x_1, ..., x_n)f_j(x_1, ..., x_n)
\eeq
where the $f_j(x_1, ..., x_n) = x_1^j + ... + x_n^j$ are computationally trivial to evaluate. With this in hand, we can rapidly evaluate Eq. \ref{eq:p_k} as
\beq
p_k(\Lambda) = \sum_{r=0}^{\min (k, N_{\rm events}-k)}p^{(r)}.
\eeq
If one wishes to calculate the $k$-expectation over the $p_k(\Lambda)$
\beq
\langle p_k(\Lambda)\rangle_k = \sum_{k=0}^{N_{\rm events}}k p_k(\Lambda), 
\eeq
it is simple enough to evaluate given all the $p_k(\Lambda)$, however it is clear that this should also equal the sum of the $\pastroi$, thinking of the $\pastroi$ as independent Bernoulli trials. We can show they are equivalent by writing down a generating polynomial for $p_k(\Lambda)$
\beq
\sum_{k=0}^{N_{\rm events}} x^kp_k(\Lambda) = \prod_{i=1}^{N_{\rm events}}\left[x \pastroi + (1-\pastroi)\right].
\eeq
Evaluating the polynomial for $x=1$ shows the $p_k(\Lambda)$ are indeed normalized, and evaluating the first derivative at $x=1$ shows the $k$-expectation is equal to the sum of the $\pastroi$.

One may be tempted to use the $k$-expectation as it has continuous support, however we caution that using only the $k$-expectation can be somewhat misleading. For some population inferences, there was very little support for 69 GW events in the $k$-expectation posterior, while there was a reasonable probability for having exactly 69 GW events in the catalog. These are different statistical statements and should not be mistaken for one another.

This kind of calculation can in principle be done for any population inference with a mixing fraction. That said, our populations are nearly disjoint and as such the posterior width on $\etabar$ is dominated by Poisson uncertainty, not uncertainty on which events in the catalog should belong to which sub-populations. For other population inferences with mixing fractions, the events may not be as easy to differentiate into sub-populations, and the uncertainty on the mixing fraction will have a larger contribution from this uncertainty. The $p_k(\Lambda)$ will have broader support and will more closely mimic the (appropriately rescaled) detectable mixing fraction posterior.

\end{document}